\documentclass[usenatbib]{mnras}

\usepackage{newtxtext,newtxmath}
\usepackage[T1]{fontenc}
\usepackage{ae,aecompl}

\usepackage{graphicx}	% Including figure files
\usepackage{xspace}
\usepackage{amsfonts}
\usepackage{textcomp}
\usepackage[usenames]{color}
\usepackage{times}
\usepackage{hyperref}

\usepackage{enumitem}
\usepackage[compatibility=false]{caption}
\usepackage{subcaption}
\usepackage{multirow}
\usepackage{tabulary}
\usepackage[para]{threeparttable}
\usepackage{array,booktabs,longtable,tabularx}
\newcolumntype{L}{>{\raggedright\arraybackslash}X}
\usepackage{ltablex}

\renewlist{tablenotes}{enumerate}{1}
\makeatletter
\setlist[tablenotes]{label=\tnote{\alph*},ref=\alph*,itemsep=\z@,topsep=\z@skip,partopsep=\z@skip,parsep=\z@,itemindent=\z@,labelindent=\tabcolsep,labelsep=.2em,leftmargin=*,align=left,before={\footnotesize}}
\makeatother

\newcommand{\msun}{\mbox{$M_\odot$}}

\newcommand{\disperse}{\mbox{{\sc \small DisPerSE}}\xspace}
\newcommand{\ssfr}{\mbox{sSFR}\xspace}
\newcommand{\vsig}{\mbox{$v/\sigma$}\xspace}
\newcommand{\mstar}{\mbox{$M_\star$}\xspace}
\newcommand{\simba}{\mbox{{\sc  Simba}}\xspace}
\newcommand{\gizmo}{\mbox{{\sc \small Gizmo}}\xspace}
\newcommand{\mufasa}{\mbox{{\sc \small Mufasa}}\xspace}
\newcommand{\hagn}{\mbox{{\sc \small Horizon-AGN}}\xspace}
\newcommand{\ill}{\mbox{{\sc \small Illustris-1}}\xspace}
\newcommand{\eagle}{\mbox{{\sc \small Eagle}}\xspace}
\newcommand{\mbII}{\mbox{{\sc \small Massive-Black II}}\xspace}
\newcommand{\gad}{\mbox{{\sc \small Gadget-3}}\xspace}
\newcommand{\grac}{\mbox{{\sc \small Grackle-3.1}}\xspace}
\newcommand{\yt}{\mbox{{\sc \small YT}}\xspace}
\newcommand{\caesar}{\mbox{{\sc \small Caesar}}\xspace}
\newcommand{\ramses}{\mbox{{\sc \small Ramses}}\xspace}
\newcommand{\arepo}{\mbox{{\sc \small Arepo}}\xspace}
\newcommand{\gadget}{\mbox{{\sc \small Gadget}}\xspace}

\definecolor{Orange}{rgb}{1.0,0.5,0.15}
\definecolor{Blue}{rgb}{0,0.08,0.65}
\definecolor{Red}{rgb}{0.65,0.08,0.05}
\definecolor{Green}{rgb}{0.15,0.45,0.25}

\title[]{And yet it flips: connecting galactic spin and the cosmic web}

\author[K. Kraljic, R.~Dav\'e, C.~Pichon]{
Katarina~Kraljic,$^{1}$\thanks{E-mail: kat@roe.ac.uk} 
Romeel~Dav\'e,$^{1,2,3}$
and Christophe~Pichon$^{4,5}$\\
$^{1}$ Institute for Astronomy, Royal Observatory, Edinburgh EH9 3HJ, United Kingdom\\
$^{2}$ University of the Western Cape, Bellville, Cape Town 7535, South Africa\\
$^{3}$ South African Astronomical Observatories, Observatory, Cape Town 7925, South Africa\\
$^{4}$ Sorbonne Universit{\'e}s, UPMC Univ Paris 6 et CNRS, UMR 7095, Institut d'Astrophysique de Paris, 98 bis bd Arago, 75014 Paris, France \\
$^{5}$ Korea Institute for Advanced Study (KIAS), 85 Hoegiro, Dongdaemun-gu, Seoul, 02455, Republic of Korea \\
}

\date{Accepted XXX. Received YYY; in original form ZZZ}

\pubyear{2019}

\begin{document}
\label{firstpage}
\pagerange{\pageref{firstpage}--\pageref{lastpage}}
\maketitle
\begin{abstract}
We study the spin alignment of galaxies and halos with respect to filaments and walls of the cosmic web, identified with \disperse, using the \simba simulation from $z=0-2$.
Massive halos' spins are oriented perpendicularly to their closest filament's axis and walls, while low mass halos tend to have their spins parallel to filaments and in the plane of walls. A similar mass-dependent spin flip is found for galaxies, albeit with a weaker signal particularly at low mass and low-$z$, suggesting that galaxies' spins retain memory of their larger-scale environment.
Low-$z$ star-forming and rotation-dominated galaxies tend to have spins parallel to nearby filaments, while quiescent and dispersion-dominated galaxies show preferentially perpendicular orientation; the star formation trend can be fully explained by the stellar mass correlation, but the morphology trend cannot.
There is a strong dependence on HI mass, such that high-HI galaxies tend to have parallel spins while low-HI galaxies are perpendicular, which persists even when matching samples in stellar mass, suggesting that HI content traces anisotropic infall more faithfully than the stellar component. 
Finally, at fixed stellar mass, the strength of spin alignments correlates with the filament's density, with parallel alignment for galaxies in high density environments. These findings are consistent with conditional tidal torque theory, and highlight a significant correlation between galactic spin and the larger scale tides that are important e.g. for interpreting weak lensing studies.
\simba allows us to rule out numerical grid locking as the cause of previously-seen low mass alignment.
\end{abstract}


%%%%%%%%%%%%%%%%%%%%%%%%%%%%%%%%%%%%%%%%%%%%

%%%%%%%%%%%%%%%%% BODY OF PAPER %%%%%%%%%%%%%%%%%

\section{Introduction}

Understanding the origin of the diversity of galaxy morphologies seen today, the so-called Hubble sequence, is one of the biggest challenges of the theory of galaxy formation.  The morphology of galaxies is intimately related to their angular momentum, which is acquired from the large-scale structure of the Universe through cosmic flows of matter, mergers, and interactions. The theory of structure formation thus suggests that galaxy morphology is partially driven by the large-scale anisotropic environment.

In the standard paradigm, the angular momentum (or spin) of proto-halos is at linear order induced by the misalignment between the inertia tensor of the proto-halo and the tidal tensor in its surroundings \citep[][see also \citeauthor{schafer09} \citeyear{schafer09}, for a review]{Hoyle49,peebles69,doroshkevich70,white84,CatelanTheuns1996,Crittenden2001}. 
At later stages, as the proto-halos decouple from cosmic expansion and collapse into virialised structures, strongly non-linear processes may impact galaxies' angular momentum distribution \citep{porciani02a,porciani02b}.  For instance, galactic outflows can redistribute angular momentum within the inner parts of galaxy halos~\citep{danovichetal12}, and increase the angular momentum of disks via wind recycling~\citep{brook11,christensen16}.

Since the cosmic web is shaped by the same gravitational tidal field responsible for the acquisition of the net angular momentum of systems forming and evolving within, a correlation between the large-scale structure and the spin of halos is directly expected from Tidal Torque Theory (TTT). 
By revisiting TTT in the context of such anisotropic environment (filaments embedded in walls), \cite{Codis2015a} showed that the constrained misalignment between the tidal and the inertia tensors in the vicinity of filament-type saddle points is able to explain the relative angular momentum distribution with respect to the cosmic web. In particular, such conditional scale-dependent tides imply 
a spin aligned with filaments for low mass halos, and a perpendicular spin orientation for more massive halos.  This conditional tidal torque theory agrees with findings seen in halos from cosmological N-body simulations \citep[e.g.][]{AragonCalvo2007,Hahn2007,Codis2012,Trowland2013,WangKang2017,GaneshaiahVeena2018}.

A key prediction of the conditional tidal torque theory is thus that the spin orientation with respect to the embedding filament flips from low to high mass galaxies.  This mass dependent flip of the spin can be understood qualitatively in the context of the dynamics of large-scale cosmic flows and mass accretion history of halos within the anisotropic large-scale structure. The first generation of halos formed in vorticity-rich filaments, and are thus expected to have their spin aligned with their embedding filament. At later stages, as the flow along filaments also shell-crosses, halos flowing towards nodes of the cosmic web convert their orbital angular momentum into a spin perpendicular to the filament axis as they merge and grow in mass \citep[e.g.][]{Codis2012,Welker2014,Codis2015a,KangWang2015,laigle2015,WangKang2017,WangKang2018}.

All these processes affect galaxy spin alignments as well, but with a caveat -- baryon specific effects (e.g. gas inflows, stellar and black hole-driven gas outflows, cooling and heating, instabilities, etc.) are additionally expected to impact their angular momentum.  As a result, the relative spins of galaxies and their host halos can show significant misalignments, depending on redshift, mass, or the central/satellite nature of the host  \citep[see e.g.][]{Tenneti2014,Velliscig2015,Chisari2017}.
Due to the complexity of the processes involved in the formation and evolution of galaxies, our understanding of the details of galaxy-spin alignment is best examined in large-scale hydrodynamical simulations capable of capturing these highly non-linear processes.

\cite{hahn10} performed the first study addressing the spin alignment of galaxies with respect to their large-scale environment by analysing a sample of $\sim$ 100 disc galaxies in the region of a large-scale filament resimulated using the "zoom-in" technique with adaptive mesh refinement (AMR) code \ramses \citep{Teyssier2002}. With the caveat of a limited statistical significance due to small number statistics, this work reported that the most massive disk galaxies at all redshifts tend to be aligned with the direction of the filament. 
Using the large-scale simulation \hagn \citep{Dubois2014} employing the same numerical technique, \cite{Codis2018} analysed galaxy spin orientation with respect to filaments and walls of the cosmic web inside of a comoving 100 Mpc $h^{-1}$ cosmological volume. 
This work extended the previous study of \cite{Dubois2014} to a full cosmic evolution down to $z=0$ and also considered an additional cosmic web environment, the walls.
It confirmed the existence of a galaxy spin transition from parallel to perpendicular with respect to the filaments' direction, and analogously with respect to walls. Overall, blue or rotation-supported galaxies were found to dominate the alignment signal at low stellar mass, while red or dispersion-dominated galaxies tend to show a preferential perpendicular alignment.
Similar conclusion regarding  galaxy mass and color dependence of the alignment signal with respect to filaments was reported by \cite{Wang2018}
analysing the \ill simulation \citep{vogelsbergeretal14}, using the moving mesh code \arepo \citep{Springel2010}.
In contrast, \cite{GaneshaiahVeena2019} recently reported a propensity of galaxies for perpendicular alignment with their host filaments at all masses with no sign of a spin transition. This work made use of the \eagle simulation
\citep{schayeetal15}, based on an updated version of the smoothed particle hydrodynamics (SPH) based galaxy formation code \gadget \citep{Springel2005}.
Similar non-detection of spin transition for galaxies was also reported by 
\cite{Krolewski2019} in another SPH based simulation \mbII \citep{Tenneti2014}.   It has been noted that the hydrodynamic methodology may play a role in spin studies, as AMR codes can suffer from ``grid locking" where disk evolution is compromised by the imposed Cartesian grid, while SPH has difficulties controlling the amount of spurious shear viscosity in rotating disks.  Hence the existence and sense of a "spin flip" in alignment between low and high mass galaxies remains controversial.

A complicated picture is also found on the  observational side. 
When studying the spin alignment of disc galaxies with respect to the filaments of the cosmic web, some groups find preferentially parallel orientation for spirals \citep[][]{Tempel2013,Tempel2013b}, Scd types \citep{Hirv2017}, or both red and blue galaxies \citep{Zhang2013}, while others report either a tendency for a perpendicular orientation for spirals \citep{Lee07,Jones2010,Zhang2015} and Sab galaxies \citep{Hirv2017},
or even no signal at all \citep{Pahwa2016,Krolewski2019}. 
There seems to be a much better agreement for elliptical/S0 galaxies, which are found to have their spin (or minor axis) perpendicular to their host filaments' direction, in line with results of shape measurements \citep[e.g.][]{OkumuraJing2009,Joachimi2011,Singh2015,Chen2019,Johnston2019}.

Studies regarding the spin alignment of galaxies within walls of the cosmic web lead to similarly contradictory conclusions.
While some works reveal a tendency for spirals to have their spins aligned with the Local Supercluster plane \citep{FlinGodlowski1986,FlinGodlowski1990,Navarro2004}, the shells of the largest SDSS and 2dFGRS\footnote{Two Degree Field Galaxy Redshift Survey \citep[2dFGRS;][]{2dfgrs2001} and the Sloan Digital Sky Survey \citep[SDSS][]{York2000}.} cosmic voids \citep{trujillo06}, or the so-called W-M  sheet in the vicinity of the Virgo Cluster and the Local Void \citep{LeeKimRey2018},
others report on the perpendicular orientation \citep{FlinGodlowski1990,Varela2012,Zhang2015} or detect no signal \citep{slosar09,Tempel2013b}. 
For elliptical galaxies, only a weak correlation has been detected between their minor axes and the normal to the sheet \citep{Tempel2013b}.  There is currently no consensus to explain the differences seen among these various observations.

The method used to trace and quantify the cosmic web may play a crucial role in these measurements.  Filament finding algorithms have a long history, with early attempts relying on the moment of inertia tensor~\citep{Dave1997}, or Morse theory~\citep{novikov06} to define the local filament direction.
In the last decade, various filament tracers have been developed \citep[see][for a comparative study of various cosmic web extraction techniques]{Libeskind2018} which 
make distinct assumptions and deal differently with the range of probed scales, and thus lead to substantial diversity in some of the extracted properties of the cosmic web. Consequently, the efficiency of these estimators may impact our ability to quantify alignment signals (Welker et al. submitted).  Hence it is important to define the cosmic web in a way that is uniform and consistent, via a method that can be equally applied to observations in order to conduct fair comparisons to models.

% ----------------------------------------------
This work studies the orientation of the spin of galaxies with respect to filaments and walls of the cosmic web, relying on a new large-scale hydrodynamical simulation \simba \citep{Dave2019}. Our focus is first on the mass dependent flip of the spin, and its evolution in the redshift range $0 \leq z \leq 2$, for which contradictory results exist. Next, the alignment signal is investigated at $z=0$ as a function of the internal properties of galaxies, namely their morphology, specific star formation rate (\ssfr), HI mass and central/satellite dichotomy, and as a function of external properties, such as galaxies' halo mass and  filaments' density.
We also extend the recent studies of the spin alignment of galaxies with respect to the walls of the cosmic web. This investigation builds on previous work in several aspects. The \simba\ simulation employs a different hydrodynamic solver (a Meshless Finite Mass, or MFM, scheme) as opposed to SPH or AMR codes used for these studies (e.g. there is no shear viscosity in MFM), which allows the evolution of an equilibrium disk for many rotation periods without numerical fragmentation or grid locking~\citep{Hopkins2015}. \simba also includes a novel implementations of feedback from Active Galactic Nuclei (AGN) and star formation that results in comparably good or better agreement with a wide range of global galaxy properties~\citep{Dave2019} compared to state-of-the-art galaxy formation simulations.
Finally, by using the same \disperse code to define the cosmic web as in \cite{Codis2018} and including orientation relative to walls, we can straightforwardly compare to those results, and eventually to observations, as we will do in future work.

The outline of this paper is as follows. Section~\ref{sec:data} presents some of the main aspects of the \simba simulation and briefly describes how the \disperse algorithm is used in order to identify filaments and walls of the cosmic web. 
Section~\ref{sec:haloSpin} investigates the alignments of spins of halos with filaments and walls together with their redshift evolution.
The results on the alignment of the spin of galaxies with respect to filaments and walls and their redshift evolution are reported in Section~\ref{sec:results_gals}. In particular, we investigate the dependence of the alignment signal on the internal properties of galaxies such as their stellar mass, star formation activity, and their HI content.
Section~\ref{sec:environment} is dedicated to the dependence of the spin-filament orientation of galaxies on environment, parametrised by the mass of their host halo, the central/satellite dichotomy and density of their host filaments.  
Finally, Section~\ref{sec:summary} concludes.
The statistical significance quantified by Kolmogorov-Smirnov tests for the Figures is given in Appendix~\ref{app:KS}. 

Throughout this paper, by log, we refer to the 10-based logarithm. 
If not stated differently, statistical errors are computed by bootstrapping, such that the errors on a given statistical quantity correspond to the standard deviation of the distribution of that quantity re-computed in 100 random samples drawn from the parent sample with replacement.

% ----------------------------------------------
\section{Virtual Universe}
\label{sec:data} 

% -----------------
\begin{figure*}
\centering\includegraphics[width=\textwidth]{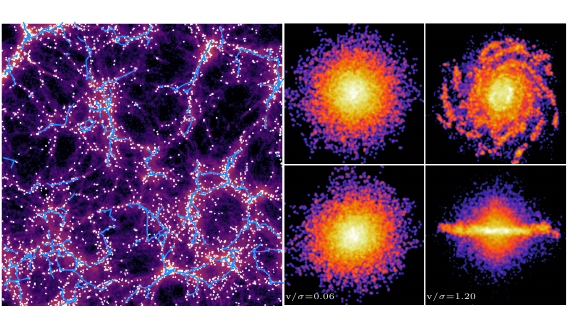}
\caption{{\sl Left:} A 2D projection of a 10 Mpc's simulation slice at $z=0$. Galaxies (white circles) are overplotted on the gas distribution. The blue lines show the filaments as extracted by the \disperse code from the galaxy distribution using the persistent threshold of 3$\sigma$. {\sl Right:} Examples of two galaxies of similar mass, $\log (\mstar/\msun) \sim 11.2$, but with very different morphology as traced by their \vsig. The galaxy on the left is  an elliptical  with low \vsig, while the galaxy on the right has a disk-dominated morphology characterised by relatively high \vsig. For both galaxies face-on and edge-on projections are shown on top and bottom panels, respectively.
}
\label{fig:simu}
\end{figure*}

We first describe our virtual universe and the analysis tools we use to trace filaments, identify galaxies and measure their physical properties.

% --------------
\subsection{The Simba simulation}
\label{sec:simba}
This work makes use of the \simba simulation \citep{Dave2019} to follow  galaxy and structure formation across cosmic time.  \simba is a new large-scale cosmological hydrodynamical simulation built on the \mufasa suite~\citep{Dave2016} that is seen to successfully reproduce many observables, such as galaxy stellar mass functions at $z=0-6$, the stellar mass-star  formation  rate  main  sequence, HI and H$_2$ galaxy gas fractions, the  mass-metallicity relation at $z=0-2$, star-forming galaxy sizes, hot gas fractions in massive halos, and galaxy dust properties at $z\sim 0$ \citep{Dave2019}. 
A full description of this simulation can be found in \cite{Dave2019}; here we summarise only some of its main features relevant to this work.

\simba was run using the Meshless Finite Mass version of the \gizmo code \citep{Hopkins2015}, a multi-method gravity plus hydrodynamics code based on \gad \citep{Springel2005}. 
The \simba run used in this work follows the evolution of 1024$^3$ dark matter particles and 1024$^3$ gas elements in a volume of (100 $h^{-1}$ Mpc$)^3$. The simulation begins at $z=249$ assuming a standard 
$\Lambda$CDM cosmology compatible with Planck Collaboration results \citep{PlanckCollaboration2016}, with $\Omega_m=0.3$, $\Omega_{\Lambda}=0.7$, $\Omega_b=0.048$, $H_{0}=68$ 
km s$^{-1}$ Mpc$^{-1}$, $\sigma_8=0.82$ and $n_s=0.97$.
The minimum gravitational softening length for this run is 0.5 comoving $h^{-1}$ kpc, the initial gas element mass is 1.82 $ \times 10^{7}$ \msun, and the dark matter particle mass resolution is 9.6 $\times 10^{7}$ \msun.

Radiative cooling and photoionisation heating are modeled using the \grac library \citep{Smith2017}, including metal cooling and non-equilibrium evolution of primordial elements.
A spatially uniform ionising background is assumed as specified by \cite{HaardtMadau2012}, modified to account for self-shielding, and the neutral hydrogen content of gas particles is modeled self-consistently. Because  significant amounts of neutral hydrogen can lie in an extended configuration beyond the star-forming region of galaxies, to assign HI to galaxies, all gas elements with HI fractions above 0.001 are considered, and assigned to the galaxy to which they are most gravitationally bound, i.e. its kinetic energy relative to the galaxy's center of mass velocity corrected for the potential energy from the galaxy at the gas element's location is minimised.

Star formation is based on the H$_2$ content of the gas, following the model used in the \mufasa simulation \citep{Dave2016}. The H$_2$ fraction computation is based on the metallicity and local column density and follows the sub-grid model of \cite{KrumholzGnedin2011}.
The  star  formation rate  is  given  by SFR = $\epsilon_{\star} \rho_\mathrm{H_2}$ / $t_{\rm dyn}$, where $\rho_{H_2}$ is the H$_2$ density, $t_{\rm dyn}$ the dynamical time and $\epsilon_{\star}$=0.02 \citep{kennicutt98} the star formation efficiency.

During the simulation, the chemical enrichment model tracks eleven elements (H,He,C,N,O,Ne,Mg,Si,S,Ca,Fe), with enrichment  tracked from Type II supernovae (SNe), Type Ia SNe, and Asymptotic Giant Branch (AGB) stars, employing the yields of \cite{Nomoto2006} for SNII, \cite{Iwamoto1999} for SNIa, and following \cite{OppenheimerDave2006} for AGB star enrichment which is based on the \citet{bruzual&charlot03} stellar population model.
Dust production and destruction is modeled on-the-fly, following \cite{McKinnon2017} by advecting it passively with the gas elements. Additionally, \simba includes dust production by condensation of metals from ejecta of Type II SNe and AGB stars, together with further growth via condensation from metals, and destruction due to sputtering, consumption by star formation, and SNe shocks. 
We will not consider dust or metal properties in this work.

Stellar feedback is modeled via decoupled two-phase galactic winds, in which 30\% of wind particles are ejected "hot" i.e.  with a temperature set by the supernova energy minus the wind kinetic energy. 
\simba also contains an implementation of metal-loaded winds. When a wind particle is launched, some metals from nearby particles are extracted in order to represent the local enrichment by the supernovae Type II driving the wind. 
Additionally, Type Ia SNe and AGB enrichment and wind heating are included, along with interstellar medium (ISM) pressurisation at a minimum level as required to resolve the Jeans mass in star-forming gas as described in \cite{Dave2016}.

Black hole growth is modelled via the torque-limited accretion model \citep{AA2017} from cold gas and Bondi accretion from hot gas. AGN feedback is modelled via kinetic bipolar outflows, with $\sim1000$~km/s winds at high Eddington rates and up to $\sim8000$~km/s jets at low Eddingtion rates, along with X-ray energy following \citet{Choi2012}. \citet{Thomas2019} showed that \simba's black hole mass and accretion rate properties relative to the galaxy properties are generally in good agreement with observations.

Halos are identified on the fly during the run using a 3D Friends-of-Friends (FoF) algorithm within \gizmo. The linking  length is taken to be 0.2 times the mean inter-particle distance. Galaxies within halos are identified using a post-processed 6-D FoF galaxy finder. Galaxies and halos are cross-matched and their properties computed using the \yt-based package \caesar.

This work makes use of outputs of the simulation at redshifts $z=2$, 1 and 0. We consider only galaxies with \mstar $\geq 10^9 \msun$,
somewhat more conservatively than in \citet{Dave2019} owing to our desire to properly resolve the angular momentum direction of the galaxy. This results in the galaxy catalogues containing 32,048 galaxies at $z=0$, 18,417 galaxies at $z=1$ and 9,674 galaxies at $z=2$.

We define the spin of a galaxy as the angular momentum computed from its stellar particles, relative to the centre of mass of the stellar component. The angular momentum (or spin) $\mathbf{L}$ of galaxies is thus computed as
\begin{equation}
    {\mathbf{L}} = \sum_{i=1}^{N_{\rm stars}} m_{i} {\mathbf{x}}_{i} \times {\mathbf{v}}_{i},
\end{equation}
where $m_{i}$, ${\mathbf{x}}_{i}$ and ${\mathbf{v}}_{i}$ and the mass, position and velocity of $i$-th stellar particle relative to the center of mass of the galaxy, respectively.

The morphology of galaxies is characterized by the kinematic ratio of their rotation to dispersion dominated velocity, \vsig. This quantity is computed from the 3D velocity distribution of stellar particles of each galaxy. 
In order to define a set of cylindrical spatial coordinates ($r$, $\theta$, $z$), the total angular momentum of stars is computed first and the $z$-axis is chosen to be oriented along the spin of galaxy.
The velocity of each stellar particle is then decomposed into cylindrical components $v_r$, $v_{\theta}$, $v_z$, and the rotational velocity of a galaxy $v$ is defined as the mean of $v_\theta$ of individual stars. The average velocity dispersion of the galaxy $\sigma^2 = (\sigma^2_r + \sigma^2_\theta + \sigma^2_z)/3$ is computed using the velocity dispersion of each velocity component $\sigma_r$, $\sigma_\theta$ and $\sigma_z$.  We note that this is not directly comparable to observational measures of \vsig, and here we simply use this quantity to separate rotation-dominated from dispersion-dominated systems.

% ----------------------
\subsection{Tracing the cosmic web}
%\label{sec:cw}

The filaments and walls of the cosmic web are extracted with the use of the publicly available code \disperse \citep{Sousbie2011,SousbiePichon2011}\footnote{http://www.iap.fr/users/sousbie/disperse/}. \disperse
is a geometric 3D ridge extractor, which identifies  cosmic web structures   with  a parameter- and scale-free
topologically motivated algorithm. It uses the notion of persistence that allows to select the retained structures on the basis of the significance of the topological connection between  pairs of critical points (maxima, saddles).

For the purposes of this work, \disperse was run on the distribution of galaxies with a 3$\sigma$ persistence threshold.
It was checked that choosing higher threshold, such as 5$\sigma$, that would select topologically more robust structures, does not alter our results, in agreement with previous works that explored the dependence of the results on the persistence threshold in more details \citep{Codis2018}.

Each galaxy is assigned its closest segment of the filaments, defined by a pair of points providing the direction of the filament for a given galaxy. The cosine of the angle between the spin of the galaxy and its closest filament, $\cos \theta$ is measured and used to assess the alignment with respect to the filamentary structure of the cosmic web. Values of $\cos \theta$ close to 1 mean that galaxy tends to have its spin aligned with the neighbouring filament, while values close to 0 mean that the spin is in the perpendicular direction with respect to the filament's axis.
Similarly, each galaxy is assigned its closest triangle, the sets of which define the tesselation of the walls (2D analogue of the sets of segments defining filaments in 1D). The direction of the wall is defined by means of the normal vector to the triangle. The cosine of the angle between the spin of the galaxy and the normal to its closest wall, $\cos \theta$ is measured and used to assess the alignment with respect to the walls. Values of $\cos \theta$ close to 0 mean that galaxy tends to have its spin aligned with the neighbouring wall, while values close to 1 mean that the spin is  perpendicular to the wall.  In practice, in order to increase the statistics of the measured signal, each galaxy is assigned two closest filaments and walls, however, considering only one closest filament and wall does not alter our results. 
In order to quantify the likelihood whether the measured alignments are consistent with being derived from a uniform distribution, a Kolmogorov-Smirnov (KS) test was performed on each distribution. The corresponding probability, $p_{\rm KS}$, for presented Figures can be found in Appendix~\ref{app:KS}.

Figure~\ref{fig:simu}, left panel, shows the cosmic web identified by \disperse (blue lines), overlaid by the galaxy population (white dots), projected within a random 10~Mpc slice from the \simba simulation.  The galaxies trace out the cosmic web as expected, with filaments and nodes clearly seen to follow the underlying gaseous cosmic web (shown as levels of purple).  \disperse generally does a good job at identifying the filaments that one would trace out by eye within that slice.  Note that the \disperse skeleton is typically continuous, but in some places it goes outside the chosen slice, so the blue line terminates: in 3D, the skeleton continues beyond this slice.

The right panel images show face-on ({\sl top}) and edge-on ({\sl bottom}) projections of the particle distribution for two massive galaxies, the left one randomly selected with low-\vsig, and the right one randomly selected among those with high \vsig.  This shows that \vsig traces morphologies as expected, in that low \vsig galaxies are spheroidal while high \vsig ones are disk-like.  We will examine galaxy spin alignments versus morphology later, where we will specifically use \vsig as a proxy  to quantify morphology.

% ------------------
\section{Halo spin alignment}
\label{sec:haloSpin}

% ----------------
\begin{figure*}
\centering\includegraphics[width=\columnwidth]{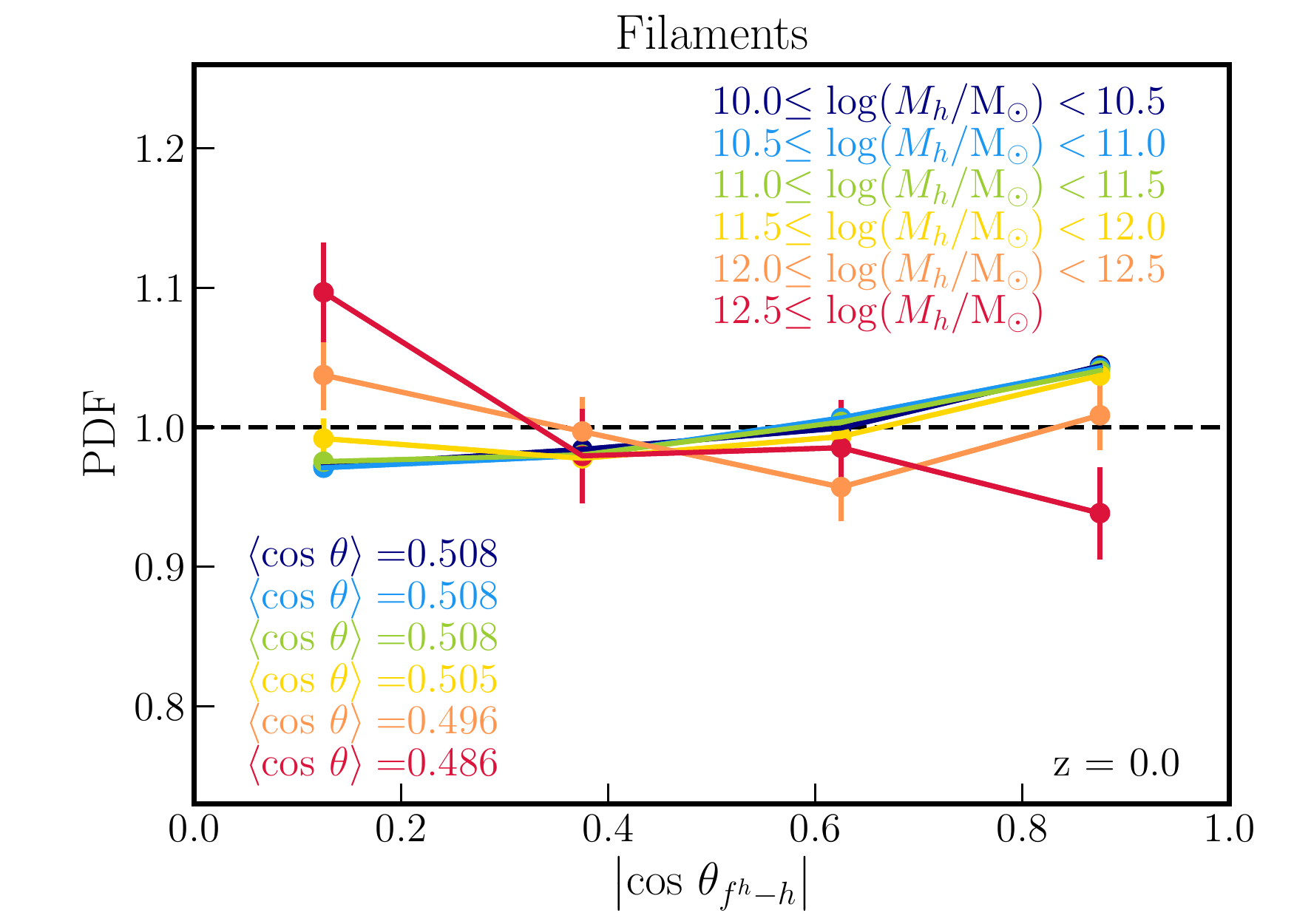}
\centering\includegraphics[width=\columnwidth]{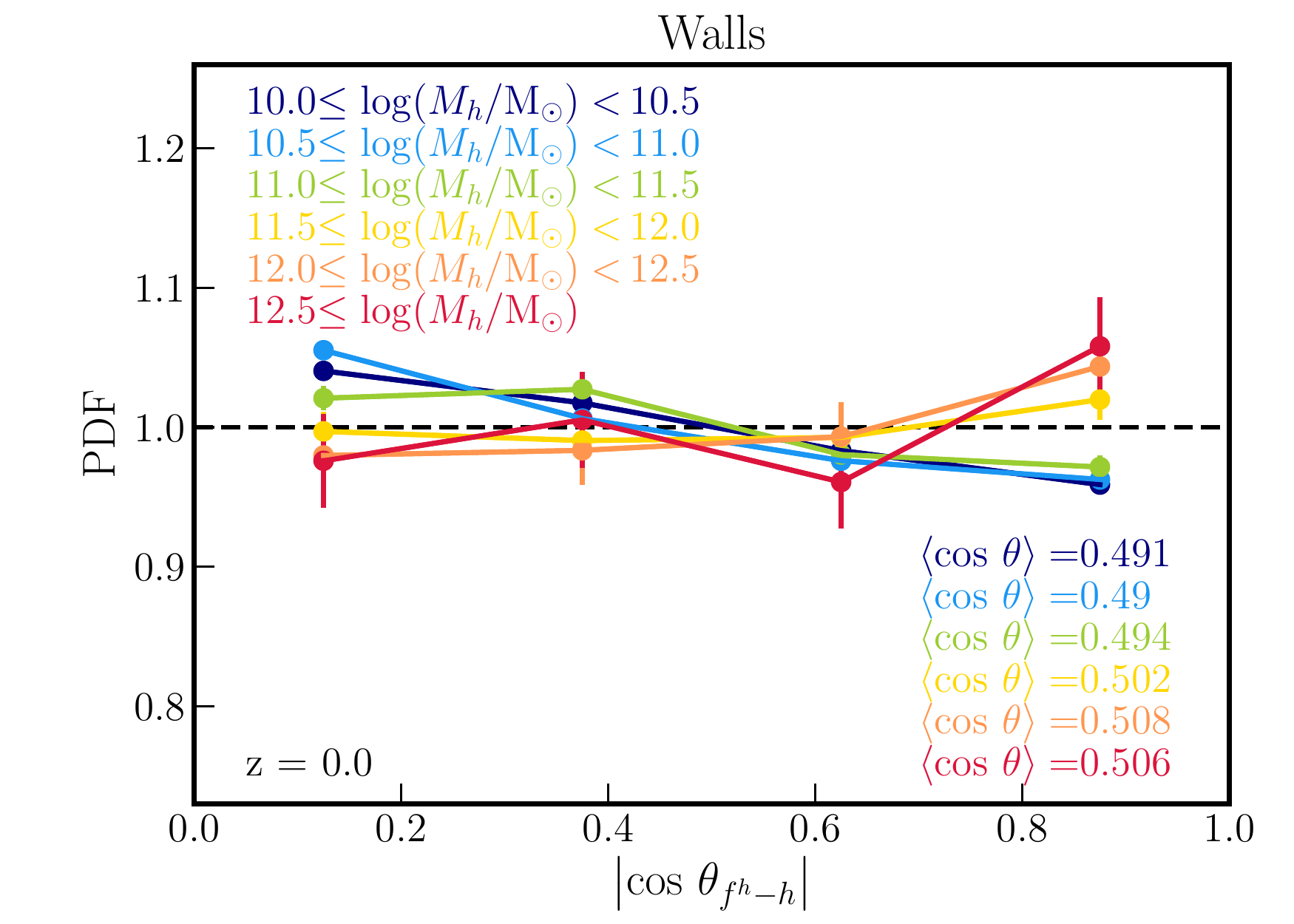}\\
\centering\includegraphics[width=\columnwidth]{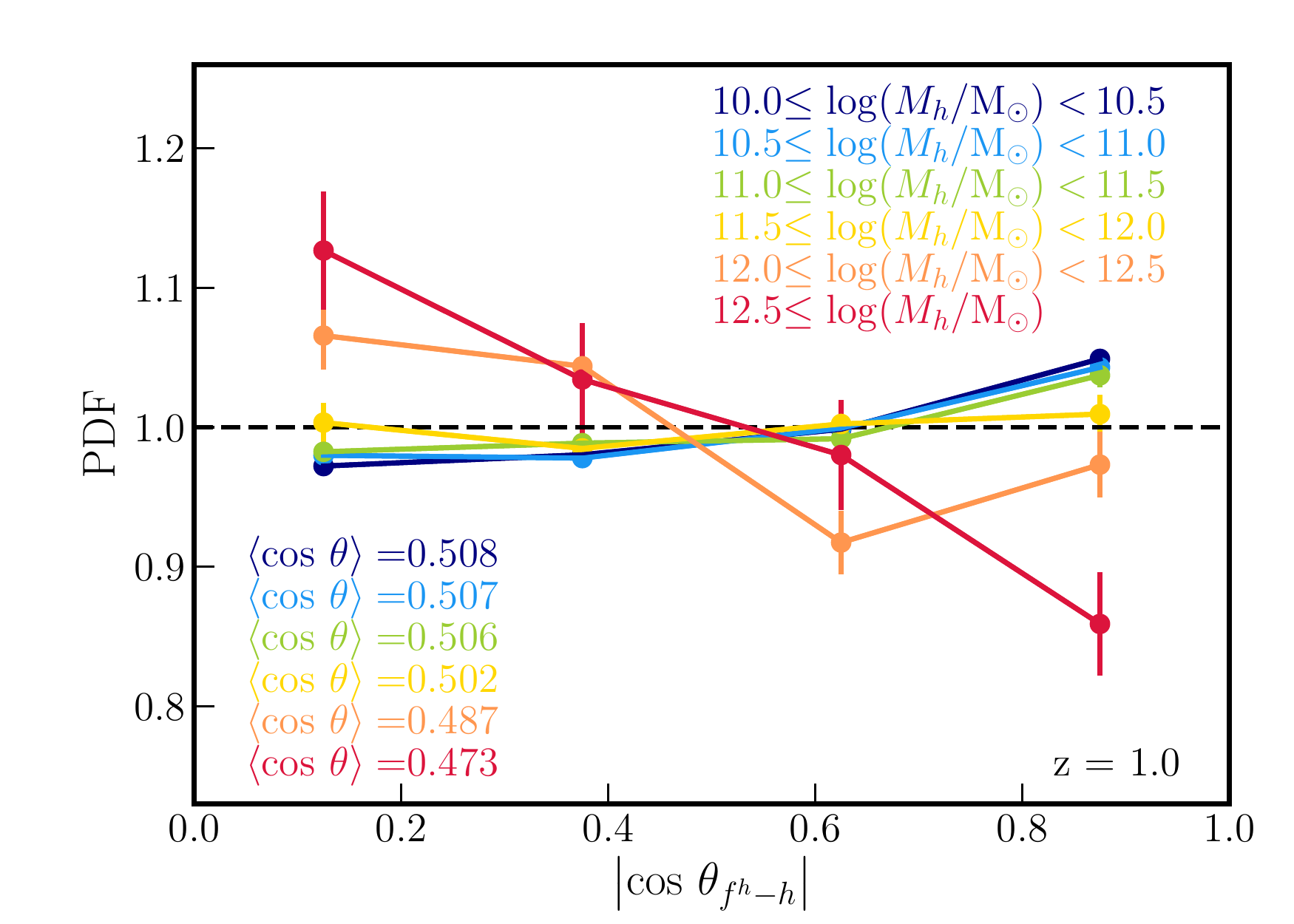}
\centering\includegraphics[width=\columnwidth]{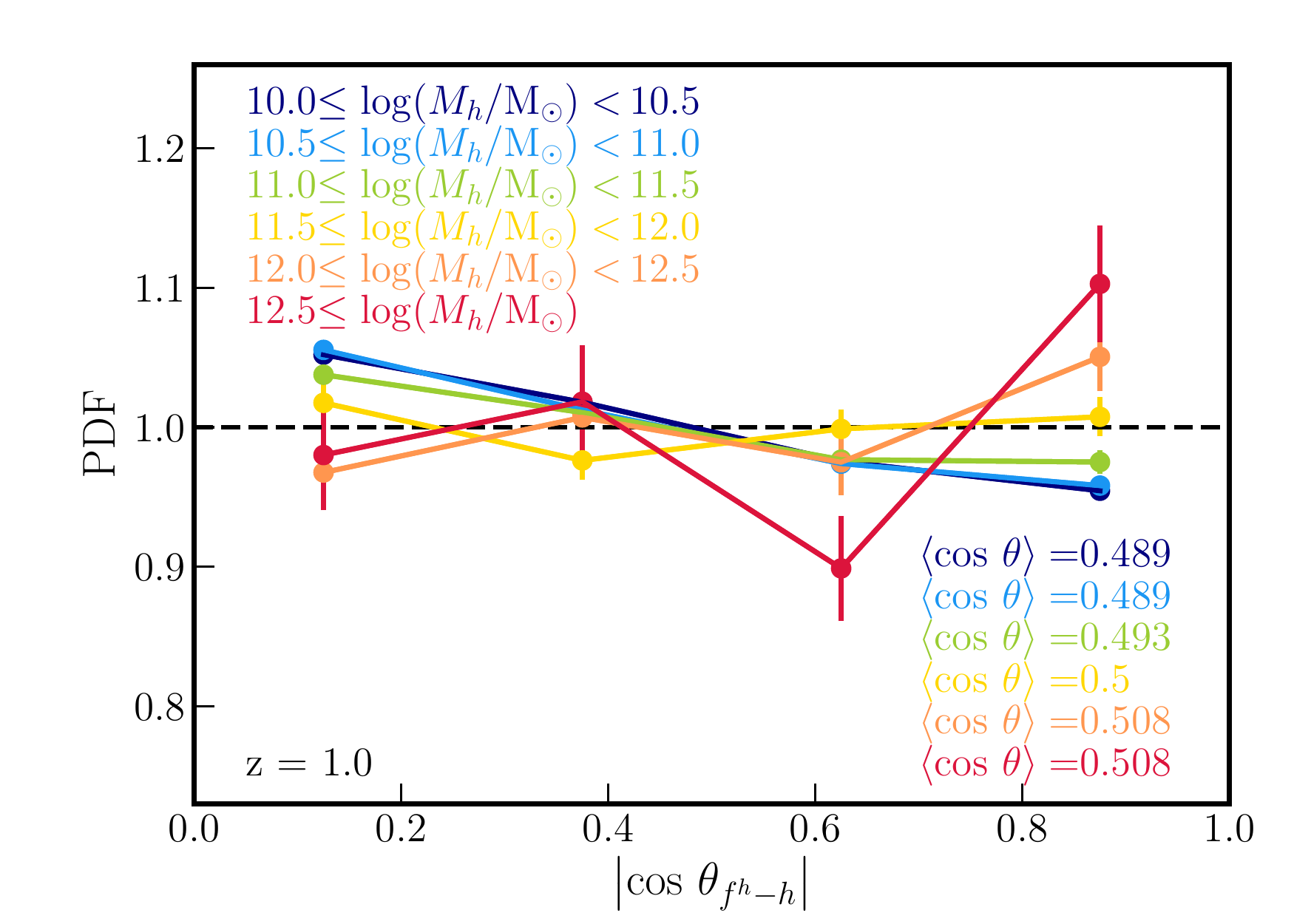}\\
\centering\includegraphics[width=\columnwidth]{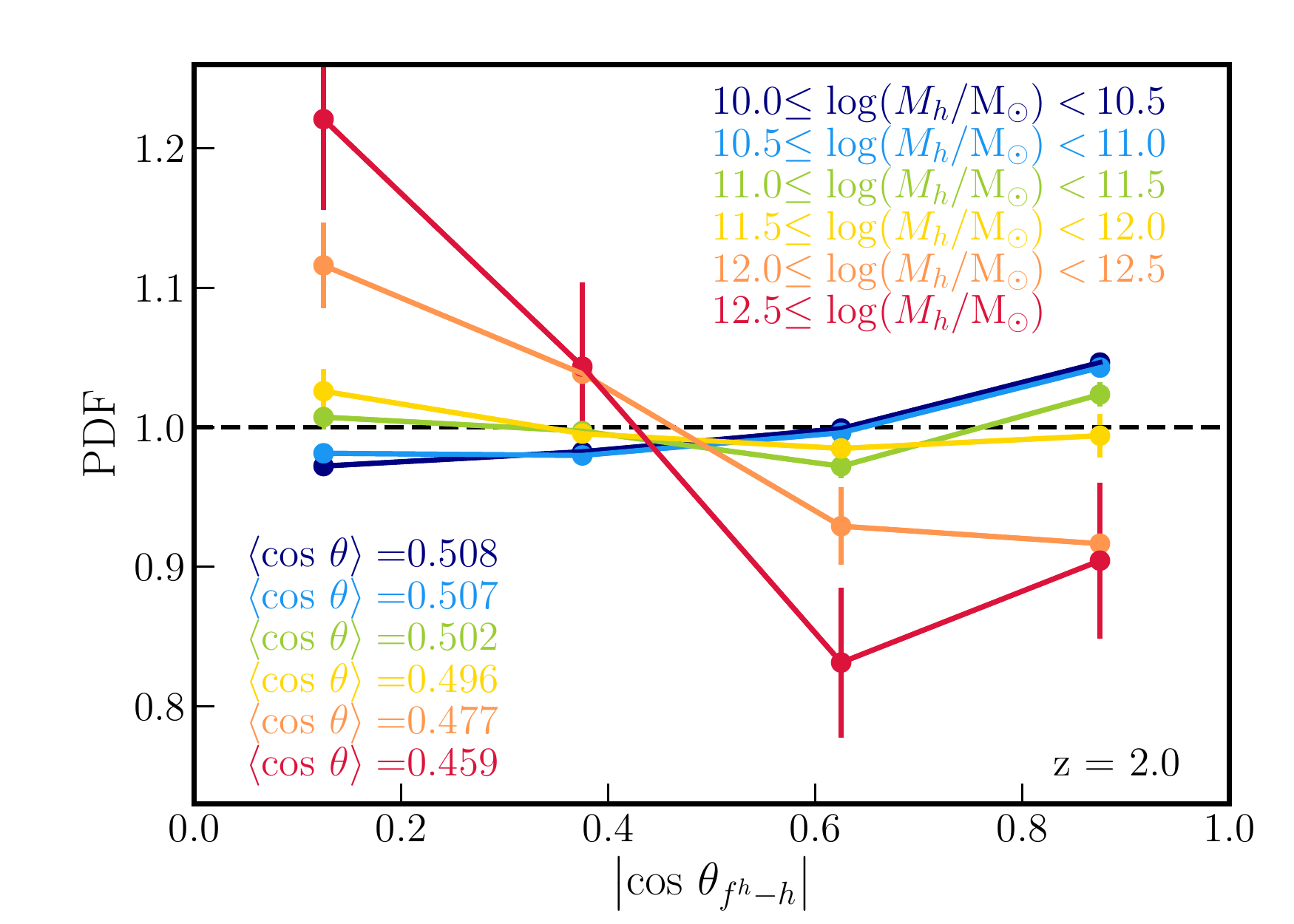}
\centering\includegraphics[width=\columnwidth]{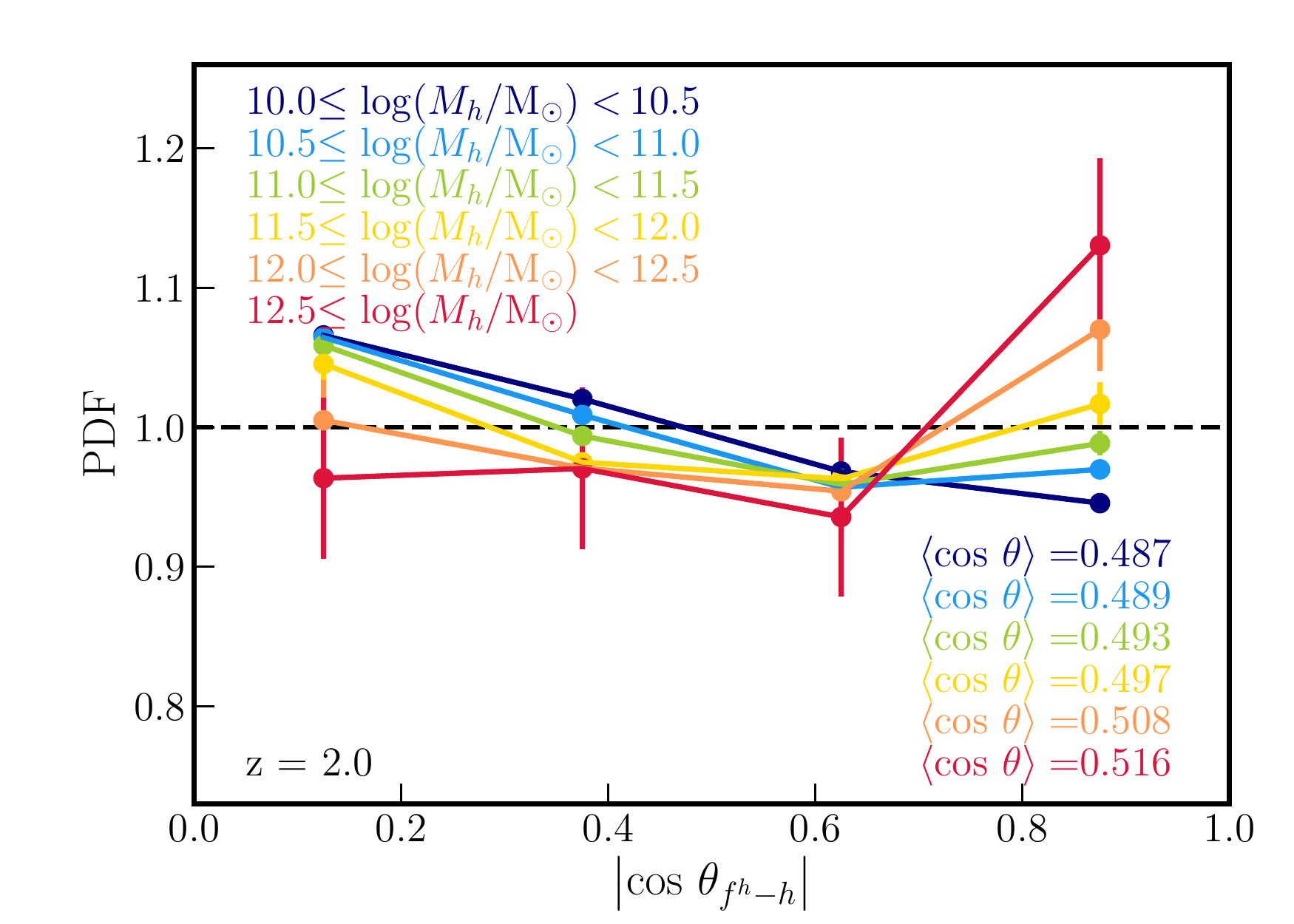}
\caption{Alignment between the spin of halos and filaments (\textsl{left}) and walls (\textsl{right}) in different halo mass bins, as labelled, at redshift $z=0$ (\textsl{top row}), 1 (\textsl{middle row}) and 2 (\textsl{bottom row}). The cosmic web is reconstructed from the distribution of halos with $\log (M_h/\msun) \geq 10.0$. The error bars represent the Poisson noise. The horizontal black dashed line represents a random distribution. Massive halos tend to have their spin perpendicular, and lower mass halos parallel to their host filaments and walls at all explored redshifts.
}
\label{fig:spin_halo_fil_halo}
\end{figure*}

Let us  start by studying the orientation of  halo spin  with respect to filaments and walls as a function of halo mass and redshift.  We consider a sample made of all halos with mass $M_h>10^{10}M_\odot$, in order to provide a more direct comparison to existing literature from both $N$-body and hydrodynamic simulations. It also sets the stage for understanding the trends seen in the spin alignments of galaxies.  Indeed, while the details of the alignment between the spin of galaxies and the direction of their host filaments are still debated, there now seem to be a consensus in the literature on the halo spin--filament alignment. The spin of massive halos is found to be preferentially  perpendicular  to  filaments' direction and walls, while at the low mass end, halos' spin tend to be aligned with their host filaments and walls, in both pure DM-only \citep[e.g.][]{AragonCalvo2007,Hahn2007,sousbie08,Codis2012,GaneshaiahVeena2018} and simulations containing baryons \citep[][]{Dubois2014,Codis2018}. 

In order to extract the cosmic web for halos, we run \disperse on the distribution of dark matter halos with a 3$\sigma$ persistence threshold.
Figure~\ref{fig:spin_halo_fil_halo} shows the probability distribution function (PDF) of the cosine of the angle between the spin of halos and the direction  of their closest filament $\cos \theta_{f^h-h}$ (\textsl{left}), and the cosine of the angle between the spin of halos and the normal  of their their closest wall $\cos \theta_{w^h-h}$ (\textsl{right}) at redshifts 0, 1 and 2, in various halo mass bins as indicated.  The mean angle within each halo mass bin is indicated on each panel.
There is a clear halo mass-dependent transition: low mass halos tend to have their spin aligned with their closest filament and wall, while at high mass, their spin tend to be perpendicular to the direction to the closest filament's axis and wall.  
The transition mass where the spins are randomly oriented is around $\sim 10^{11.5}-10^{12.5} \msun$.  
This result from \simba  is in general agreement with values reported in the literature, typically between $5 \times 10^{11} h^{-1} \msun$ and $5 \times 10^{12} h^{-1} \msun$ \citep{Codis2012}.

The amplitude of the signal, particularly the alignments with filaments, increases with increasing redshift.  High mass halos at high redshift are quite likely to be perpendicular to their host filament.  Low mass halos show less variation with redshift \citep[see also][for a similar trend for the shape of dark halos]{Chisari2017}.
The transition mass also varies with redshift,  like the typical mass collapsing at that redshift, the so-called mass of non-linearity, which, as shown by \cite{Codis2012}, increases with decreasing redshift  as $1/(1+z)^{2.5}$ on those scales.  

This quantitative agreement with previous work are expected, because baryonic effects do not play a major role in altering the spins of the halos. However, moving to smaller scales where baryonic processes become more important, it is less obvious how the spin of the halos relates to the spin of the stellar component of the galaxies, and in turn or independently how galaxy spins align with nearby filaments. This is what we examine next.

% -------------------------------------------
\section{Galaxy spin alignment}
\label{sec:results_gals}

\begin{figure*}
\centering\includegraphics[width=\columnwidth]{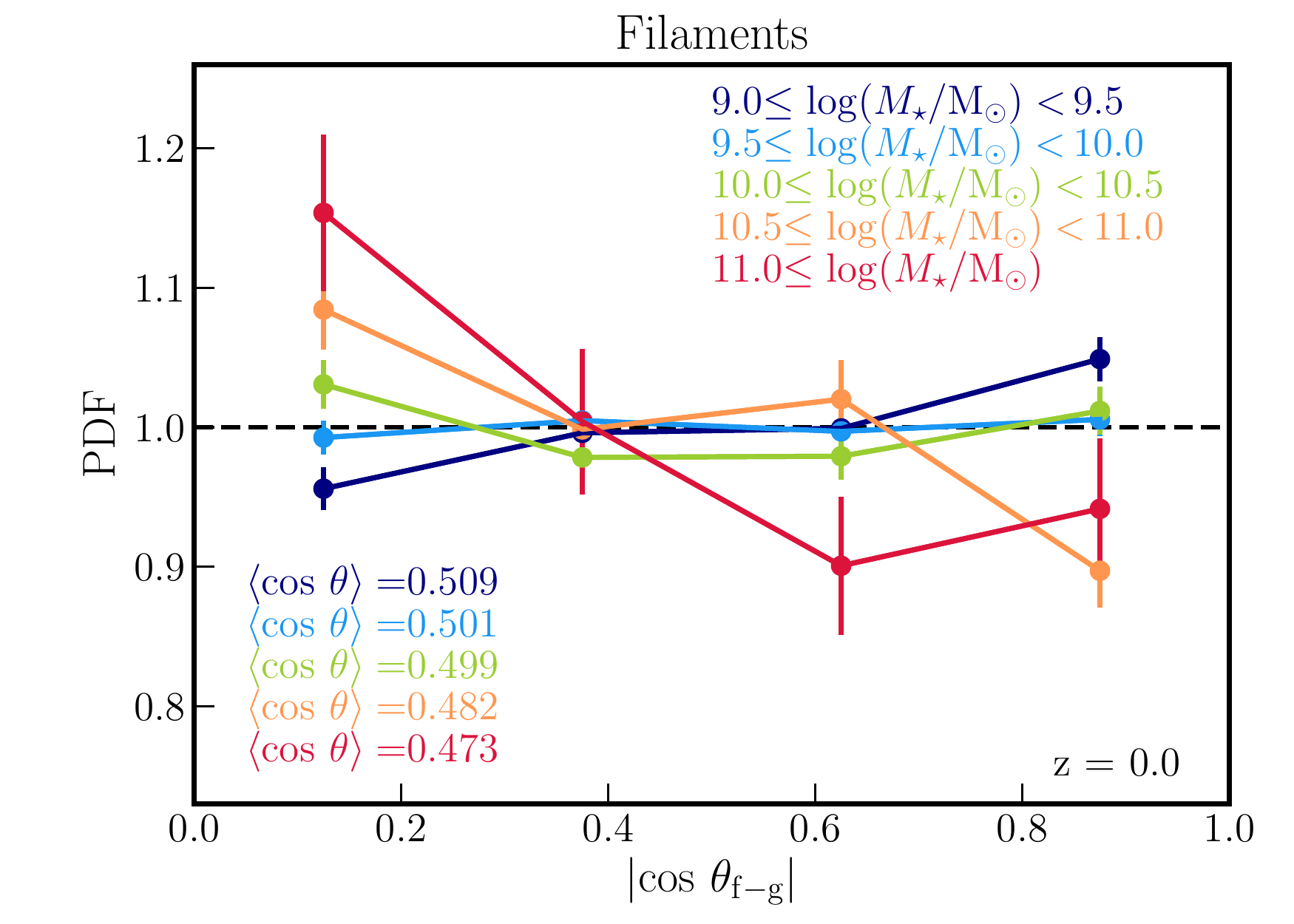}
\centering\includegraphics[width=\columnwidth]{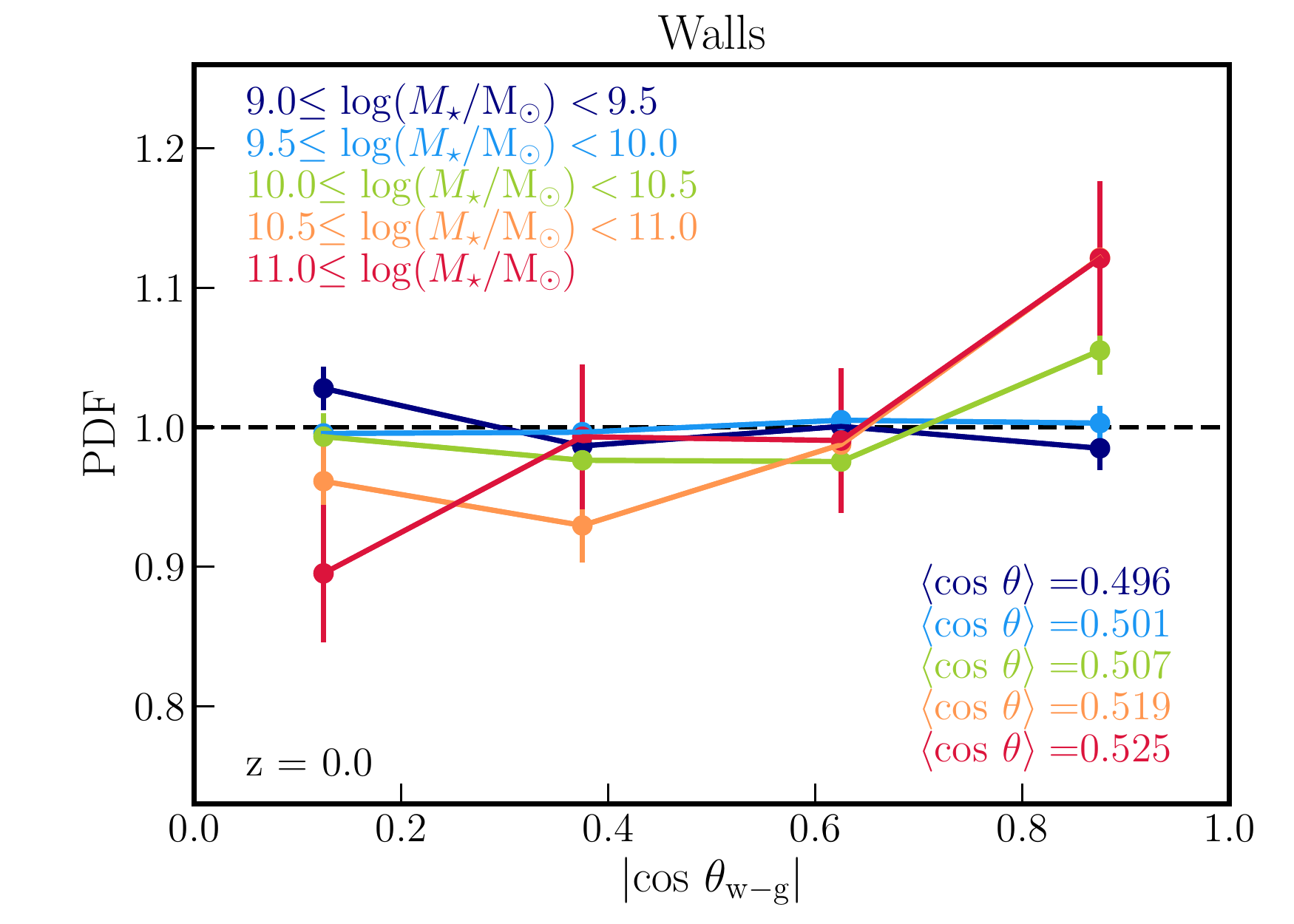}\\
\centering\includegraphics[width=\columnwidth]{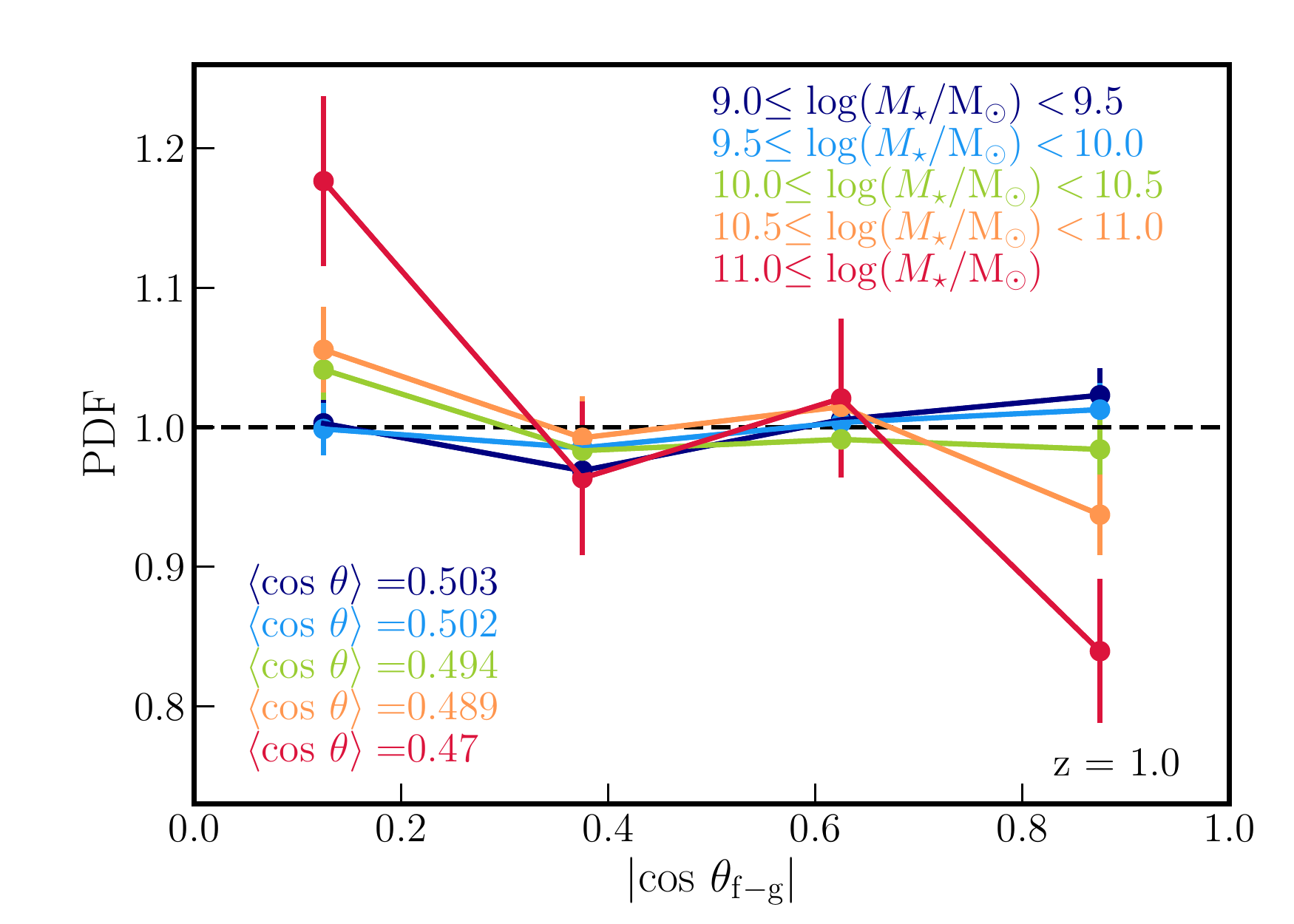}
\centering\includegraphics[width=\columnwidth]{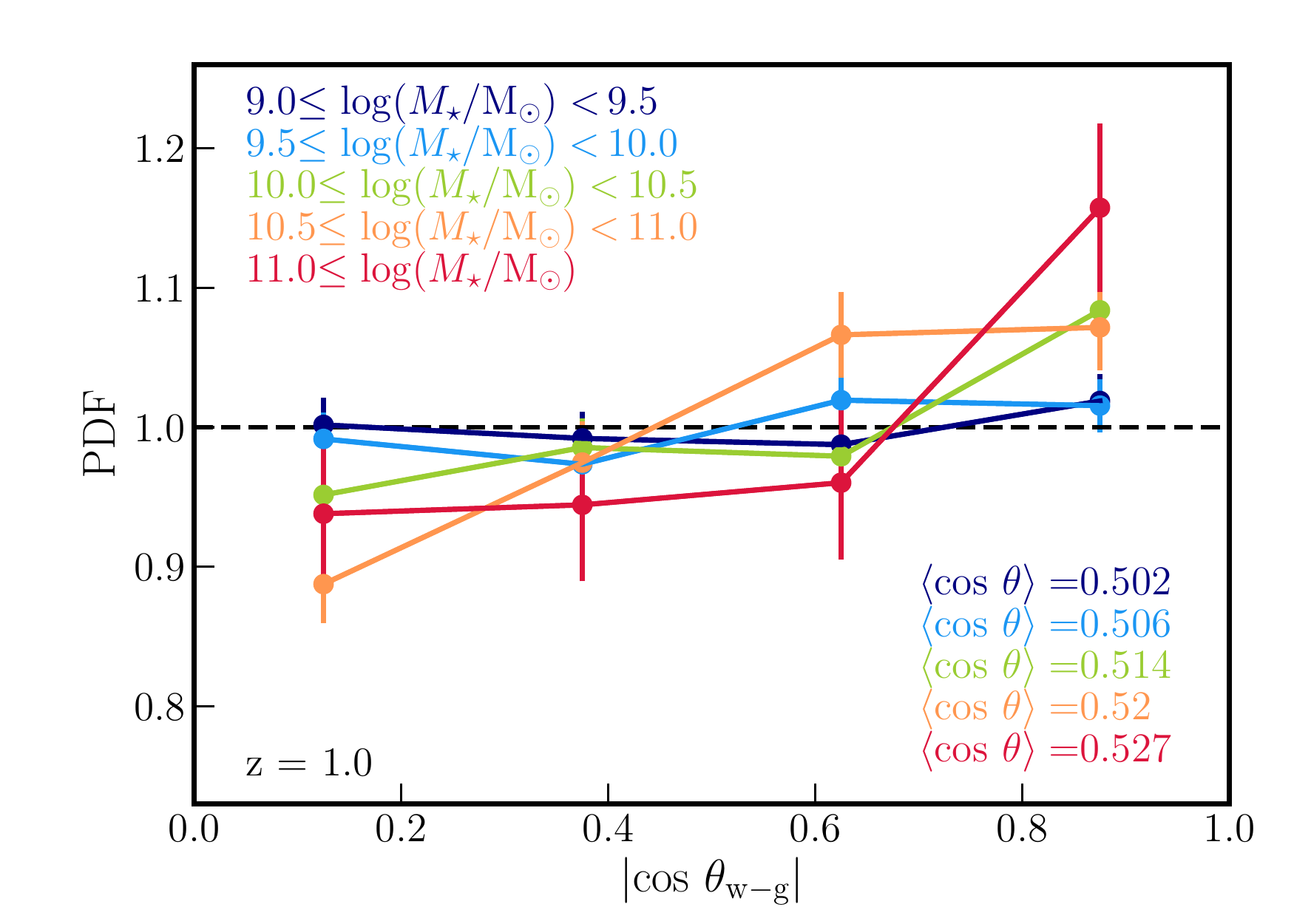}
\centering\includegraphics[width=\columnwidth]{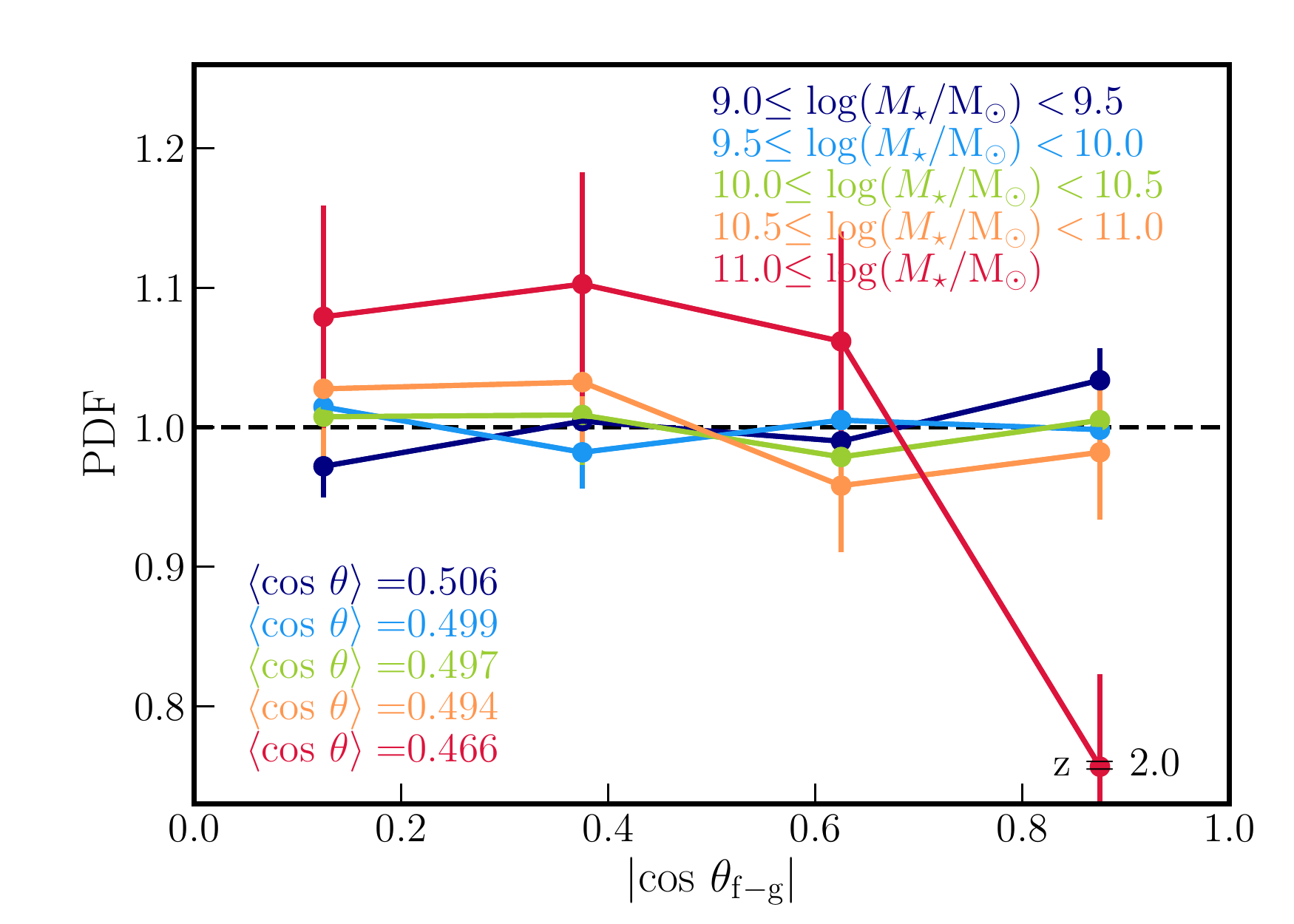}
\centering\includegraphics[width=\columnwidth]{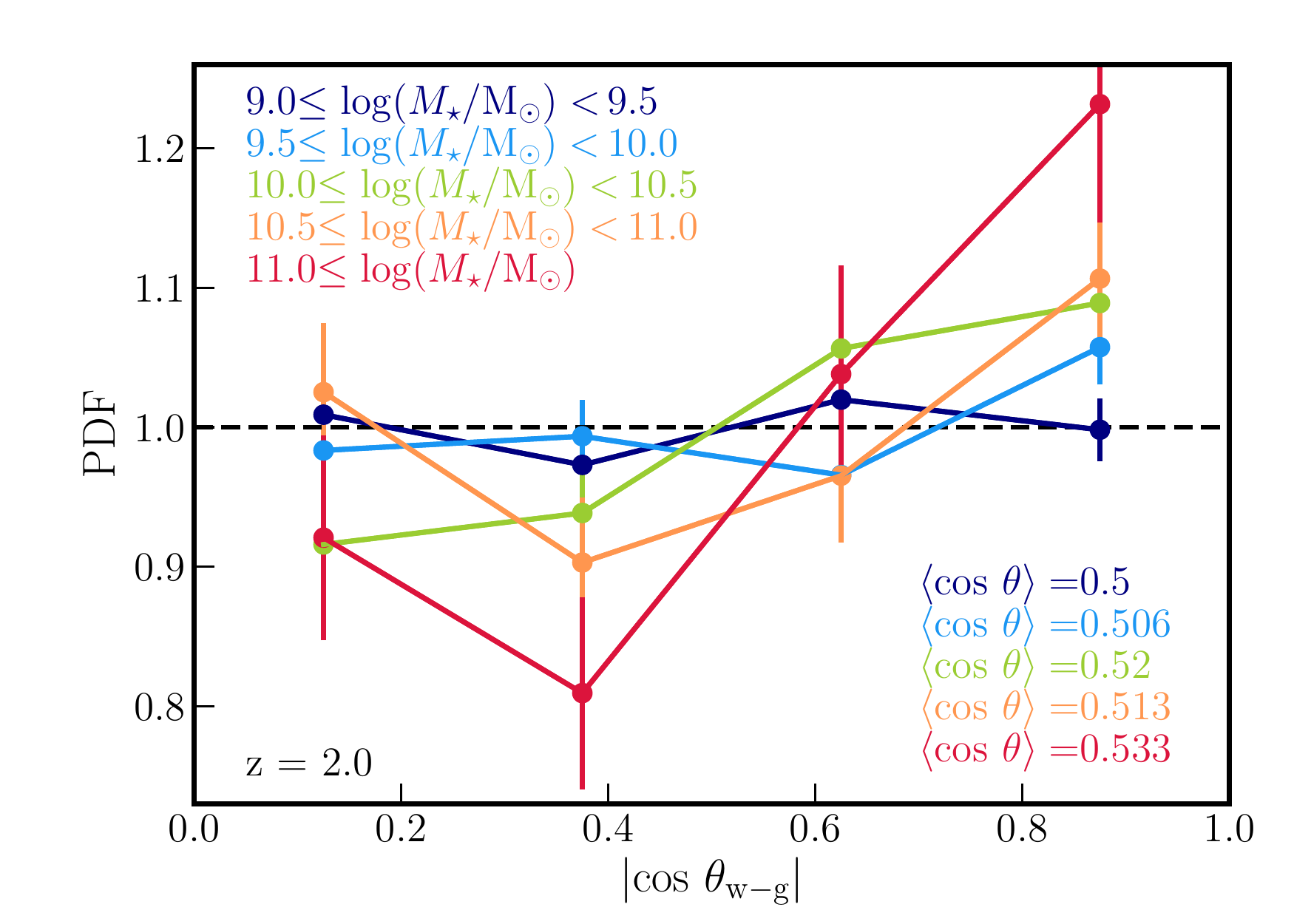}
\caption{Alignment between the spin of galaxies and their closest filaments (\textsl{left column}) and walls (\textsl{right column}) in different stellar mass bins, as labelled, at redshifts $z=0$ (\textsl{top row}), $z=1$ (\textsl{middle row}) and $z=2$ (\textsl{bottom row}). The error bars represent the Poisson noise. The horizontal black dashed line represents a random distribution. Massive galaxies tend to have their spin  perpendicular  to their host filaments and walls. At low mass, the spin of galaxies is preferentially aligned parallel to filaments and walls. This stellar mass dependent flip of the spin is detected at all explored redshifts. 
}
\label{fig:spin_fil}
\end{figure*}

\simba forms and evolves galaxies, so we can study the spin alignments of the galaxies directly with respect to the filaments and walls of the cosmic web.  For now we focus on the spin of the stellar component; we will discuss the gas component spin later.  Also, since galaxies have many other properties, we can examine the dependence of the spin alignment signal on various internal properties such as their stellar mass, their star formation activity as quantified by \ssfr, their morphology as quantified by \vsig, their gas content as quantified by their neutral hydrogen (HI) mass, and their host halo mass.  Furthermore, we will examine the dependence of the spin alignment on environmental factors, by studying spin alignments of galaxies separated into centrals and satellites and relative to the density of the nearest filament.

% -------------------------------------------
\subsection{Stellar mass dependence}
\label{subsec:mstar}

Let us start by quantifying the alignment of the spin of galaxies with respect to the filaments and walls as a function of galaxy stellar mass at different redshifts.  In \simba, as in most cosmological models, the halo mass and galaxy mass are fairly tightly correlated, so in a simple model where the baryonic angular momentum reflects some fraction of that of the dark matter \citep[as often assumed in semi-analytic models, for instance, see e.g.][and references therein]{Benson2010}, one would expect broadly similar trends as what we saw for halos.

Figure~\ref{fig:spin_fil} shows the PDF of the cosine of the angle between the spin of galaxies and the direction vectors of their closest filament $\cos \theta_{f-g}$ (\textsl{left}), and the cosine of the angle between the spin of galaxies and the normal vectors of their their closest wall $\cos \theta_{w-g}$ (\textsl{right}) at redshifts $z=0$, 1 and 2, in different stellar mass bins.  This is analogous to Figure~\ref{fig:spin_halo_fil_halo} for galaxies.

\begin{figure}
\centering\includegraphics[width=\columnwidth]{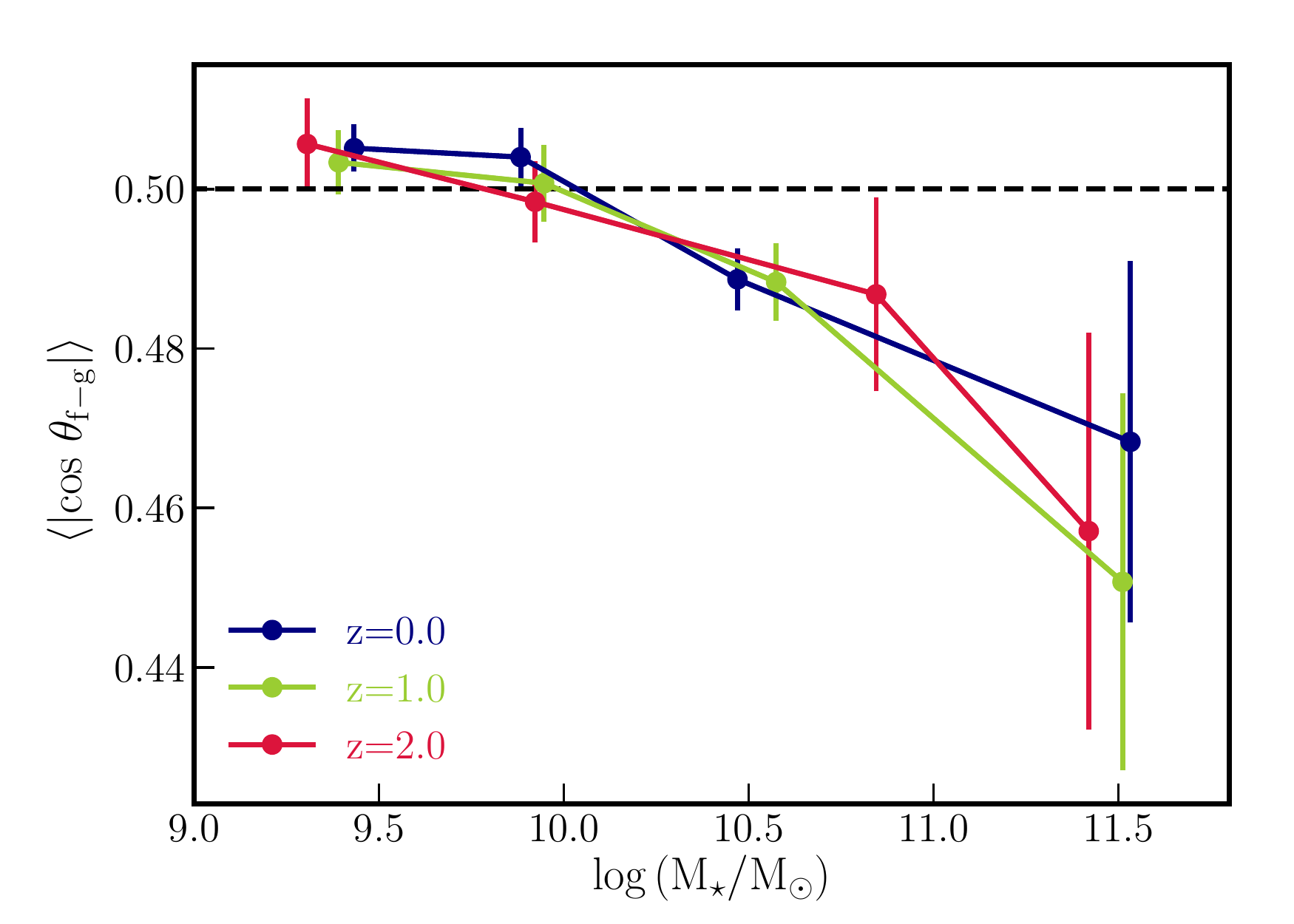}
\caption{Mean alignment between the spin of galaxies and their host filaments as a function of \mstar at redshifts $z=$2, 1 and 0, as labelled. The stellar mass dependent flip of the spin (from parallel to filament at low mass to orthogonal at high mass) is detected at all explored redshifts. 
}
\label{fig:transMstar_z}
\end{figure}

Similarly to what was seen for halos, at all redshifts there exists a stellar mass dependent transition from the parallel to perpendicular alignment, such that low mass galaxies tend to have their spin aligned with their closest filament, while high mass galaxies their spin tend to be in the perpendicular direction to the closest filament. The transition mass between these two regimes is $M_*\approx 10^{10-10.5}M_\odot$, which is approximately the stellar mass corresponding to the transition halo mass of $10^{11.5-12.5}M_\odot$.
This is more concisely shown on Figure~\ref{fig:transMstar_z}, displaying redshift and mass evolution of the mean cosine of the angle between the spin of galaxies and the direction vectors of their closest filament $\cos \theta_{f-g}$. Indeed, the transition mass at redshift 0 is roughly $10^{10.1 \pm 0.5} \msun$ 
A lack of statistics does not allow us to detect a significant evolution of the transition mass with redshift.

% ----------------

Likewise, there is a clear mass dependence of the alignment with respect to walls at all redshifts.  Low mass galaxies tend to have their spins perpendicular to the normal  of the walls, meaning that their spin lies in the plane of the wall.  High mass galaxies, conversely, have their spins preferentially aligned with the normal vector of the walls, therefore perpendicular to the plane of the walls.  The transition mass between these regimes is similar to that for the filaments.

While we do not have the ability to carry out a resolution convergence study with our current suite of simulations, we note that the spin-filament alignments were shown not to be strongly affected by the resolution in \hagn which is a comparably large simulation \citep{Codis2018} with similar resolution.  We have also tested the impact of boundary effects of a catalogue\footnote{In practice, we tested whether running \disperse with and without the option of a periodic box has an impact on the obtained results.} and found our results to be robust, which is encouraging 
in anticipation of measurements to be carried in bounded observational data sets. 

In summary, \simba produces a subtle but statistically significant trend of galaxy spin alignment with nearby cosmic filaments and walls.  The trend is mass-dependent, with low-mass galaxies having spins parallel to filaments and in the plane of walls, while high mass galaxies have spins perpendicular to filaments and the plane of walls.  The transition mass between these regimes where spins are randomly aligned is $\sim 10^{10} \msun$.  These results do not show a strong trend with redshift; in particular we do not see evidence for a stronger trend at high-$z$ as we did for halos.  Nonetheless, the overall trends generally follow that seen for halos, showing that at least statistically, the galaxies seem to follow the spin alignment behaviour of their host halos.

% -------------------------------------------
\subsection{Morphology dependence}
\label{subsec:vsig}

Galaxy properties are globally dependent on stellar mass, with low mass galaxies typically being star-forming, rotation-dominated, and (cold) gas-rich, while higher mass galaxies are the converse. Yet even
at fixed stellar mass, galaxies display some diversity in these properties. 
Thus properties such as spin alignments may have a secondary dependence when split up by other properties besides stellar mass \citep{Codis2018}.  In this section we consider spin alignments of galaxies in various mass bins when further subdividing by galaxy morphology, which we quantify by its proxy \vsig into rotation-dominated ($\vsig >0.4$) and dispersion-dominated ($\vsig <0.4$) systems, with the demarcation chosen close to the median \vsig. With this, we can examine which morphological class of galaxies is responsible for driving the trends we see with \mstar in the previous section.

% ---------------
\begin{figure}
\centering\includegraphics[width=\columnwidth]{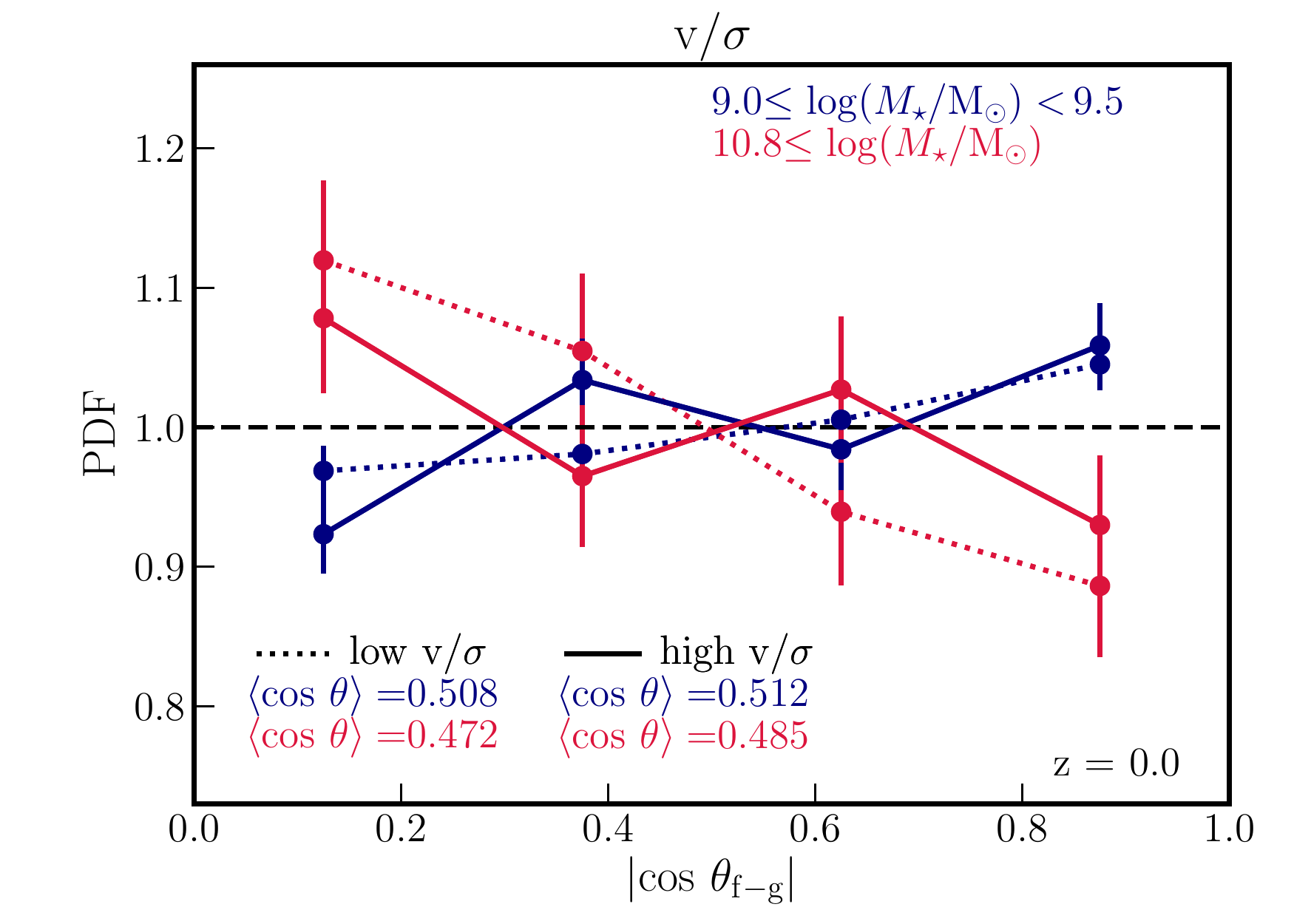}
\caption{Alignment between the spin of galaxies and their closest filaments in the lowest and highest stellar mass bins, as labelled, at redshift $z=0$ for galaxies with low (\textsl{dotted lines}) and high (\textsl{solid lines}) \vsig, corresponding to $\vsig < 0.4$ and $\vsig > 0.4$, respectively (see Table~\ref{tab:vsig} for all \mstar bins). 
 The error bars represent the Poisson noise. The horizontal black dashed line represents a random distribution. The stellar mass dependent flip of the spin seen for the entire population of galaxies is detected for both the low and high \vsig populations.
There is a hint for a tendency of galaxies with higher \vsig to dominate the parallel alignment signal at low stellar mass, while the perpendicular alignment signal tends to be dominated by galaxies with low \vsig.
}
\label{fig:spin_fil_vsig}
\end{figure}

Figure~\ref{fig:spin_fil_vsig} shows the resulting PDF of the cosines in high and low stellar mass bins for \simba galaxies at $z=0$, for galaxies with low (\textsl{dotted lines}) and high (\textsl{solid lines}) values of \vsig, respectively.  The intermediate mass alignments lie in between these extreme, and are mostly consistent with no alignment signal, so for clarity we do not show them.

The global trends are qualitatively similar for the rotation- and dispersion-dominated systems.  Both show that at low mass, galaxy spins are aligned with filaments, while at high mass they are preferentially perpendicularly aligned.  The strength of the trend is somewhat stronger in the high mass dispersion-dominated systems; in this mass bin, the rotation-dominated systems show a very weak trend.  Hence the tendency for perpendicular alignment in massive galaxies appears to be driven by the dispersion-dominated systems.

Another depiction of this trend is shown in Figure~\ref{fig:residuals_int}, upper left panel (a).  Here, we show the mean alignment as a function of \vsig, over all galaxies.  The red line shows the mean for all galaxies, and the dotted lines split these into centrals and satellites (discussed later).  The only clear alignment signal happens for high-\vsig systems, which are aligned parallel to the filaments.  The perpendicular alignment of low-\vsig systems is not evident when averaging over all galaxies.

In order to separate out the trend purely owing to morphology as opposed to that owing to a correlation between morphology and stellar mass, we examine the residuals in spin alignment versus morphology at a fixed \mstar.  This is shown in the lower left panel (d) of Figure~\ref{fig:residuals_int}.
As clearly visible in the leftmost panels (a) and (d), which show the mean $\cos \theta_{f-g}$ as a function of \vsig and its residuals at fixed \mstar, the parallel alignment at high \vsig is not simply an effect of their typically low \mstar; morphology provides an additional driver in highly rotation-dominated systems that is not accounted for purely by the trend with \mstar.  However, for the majority of galaxies with $\vsig\la 1$, any existing trend is consistent with being purely driven by \mstar.

In summary, the parallel alignment signal with nearby filaments is driven by rotation-dominated galaxies, a trend that persists even after accounting for the underlying dependence of \vsig on \mstar. This is consistent with the idea that recent cosmological accretion drives galaxies to higher \vsig, and tends to occur with an angular momentum parallel to filaments, as expected from conditional tidal torque theory, and highlighted by \cite{Welker2014} for galaxies.

% ---------------
\begin{figure}
\centering\includegraphics[width=\columnwidth]{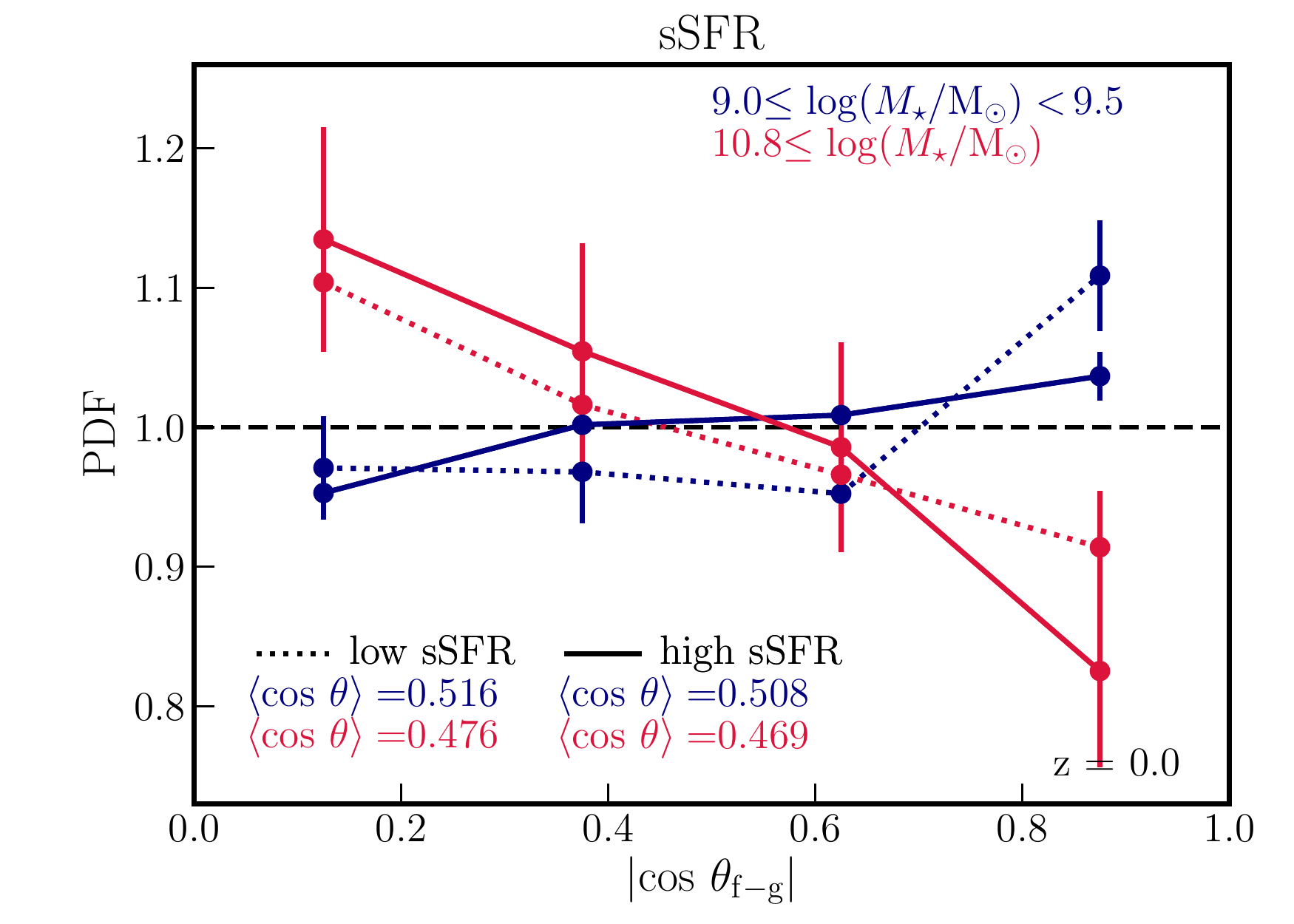}
\caption{Alignment between the spin of galaxies and their closest filaments in the lowest and highest stellar mass bins, as labelled, at redshift $z=0$ for quiescent galaxies (low \ssfr, \textsl{dotted lines}) and star-forming galaxies (high \ssfr, \textsl{solid lines}) based on the cut at $\log(\ssfr/\rm{yr}^{-1}) = -11$ (see Table~\ref{tab:ssfr} for all \mstar bins). The error bars represent the Poisson noise. The horizontal black dashed line represents a random distribution.
Both galaxy populations show a stellar mass dependent orientation of their spin with respect to filaments seen for the full galaxy sample with high (low) mass star-forming or quiescent galaxies having their spin preferentially perpendicular (parallel) to filaments. 
}
\label{fig:spin_fil_ssfr}
\end{figure}

% -----------
\subsection{Star formation rate dependence}
\label{subsec:ssfr}
% -------------

Analogously, we can examine the spin alignment signal when subdividing galaxies by their star formation activity.  We split galaxies into star-forming (blue) and quiescent (red) based on a cut in their \ssfr at $\log(\ssfr/\rm{yr}^{-1})= -11$, which is a canonical value for selecting quiescent galaxies. We note that \simba produces a quiescent galaxy fraction in quite good agreement with observations~\citep{Dave2019}.

Figure~\ref{fig:spin_fil_ssfr} shows the PDF of the cosine of the angle between the spin of galaxies and the direction vectors of their closest filament $\cos \theta_{f-g}$ for quiescent (\textsl{dotted lines}) and star-forming (\textsl{solid lines}) galaxies at redshift $z=0$, in different stellar mass bins.  As before, we only show the extreme \mstar bins, as the intermediate mass bins show essentially no alignment signal.

Once again, the flip of the spin from low to high masses is seen, regardless of their star formation activity. Both quiescent and star-forming low mass galaxies tend to have their spin aligned with the neighbouring filaments, while high mass ones have their spin preferentially in the perpendicular direction.
This suggests that the star formation activity of galaxies does not have a major impact on the alignment of galaxies.  In detail, the trend for star-forming galaxies (right panel) at the highest masses is stronger than for quiescent ones.

Figure~\ref{fig:residuals_int} (panel b) presents this in a different way, as the mean alignment angle as a function of sSFR.  This shows that there is a \ssfr-dependent flip when the entire galaxy population is considered. The high mass galaxies that dominate the low \ssfr end of the distribution are responsible for the perpendicular orientation of spin with respect filaments, while star-forming galaxies (high values of \ssfr) dominate at low masses where the spins tend to be parallel with their host filament. In fact, this trend is entirely driven by the \mstar dependence.
This is evident from Figure~\ref{fig:residuals_int} (panel e), which shows the residuals of $\cos \theta_{f-g}$ at fixed \mstar, and demonstrates that even the small alignment signal disappears once \mstar is fixed.  

Hence star formation activity provides no discernible perturbation to the alignment trend over that expected from \mstar alone.  This is interesting because the alignment trend for highly rotation-dominated systems does not appear to translate simply into a similar dependence for high-sSFR galaxies.  It appears morphology is more closely connected to spin than star formation activity in rotation-dominated systems.

% -------------------------------------------
\subsection{HI mass dependence}
\label{subsec:HI}

% ---------------
\begin{figure}
\centering\includegraphics[width=\columnwidth]{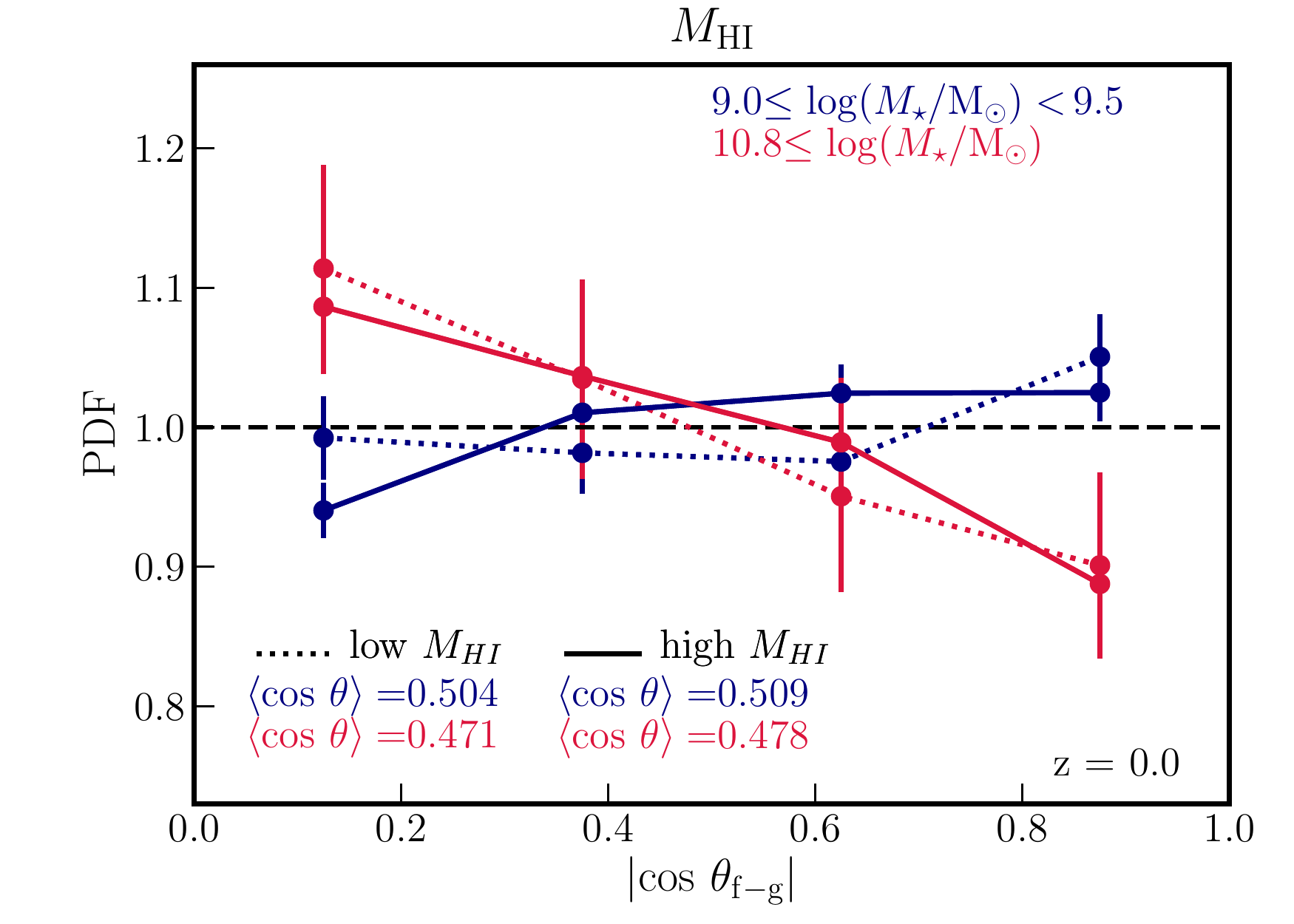}
\caption{Alignment between the spin of galaxies and their closest filaments in the lowest and highest stellar mass bins, as labelled, at redshift $z=0$ for galaxies with low (\textsl{dashed lines}) and high (\textsl{solid lines}) $M_{\rm HI}$, respectively, corresponding to split at $\log (M_{\rm HI}/\msun) = 9.17$ (see Table~\ref{tab:HI} for all \mstar bins). The error bars represent the Poisson noise. The horizontal black dashed line represents a random distribution. Stellar mass dependent flip of the spin seen for the entire population of galaxies is detected regardless of HI content of galaxies.
}
\label{fig:spin_fil_mHI}
\end{figure}

% -------------
\begin{figure*}
\centering\includegraphics[width=\textwidth]{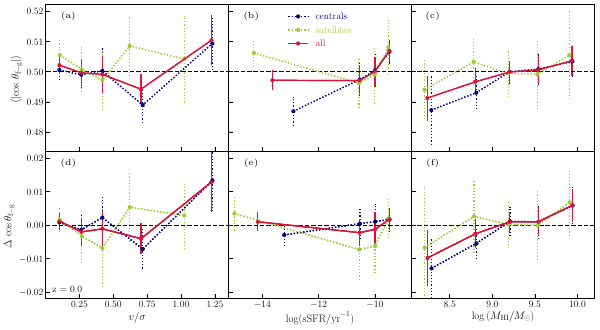}
\caption{\textsl{Top row:} Mean alignment between the spin of galaxies and filaments as a function of \vsig (\textsl{left}), \ssfr (\textsl{middle}) and HI mass (\textsl{right}) of galaxies at $z=0$. 
Galaxies with high \vsig, \ssfr and HI mass tend to align their spin with the filament's axis. 
A clear perpendicular orientation of the galaxies' spin with respect to their host filament is seen for galaxies with low HI content and for quenched centrals (with $\log(\ssfr/\rm{yr}^{-1}) \leq -11$). 
\textsl{Bottom row:} Residuals of the cosine of the angle between the spin of galaxies and the direction vectors of their closest filament $\cos \theta_{f-g}$ at fixed \mstar. Parallel spin-filament orientation of galaxies is clearly dominated by disk-dominated galaxies (with high \vsig), while all trends seen as a function of \ssfr are entirely driven by \mstar. The strongest residuals are found for HI mass, with the parallel alignment being dominated by galaxies with high HI content, while galaxies with low HI mass are driving the perpendicular orientation.
}
\label{fig:residuals_int}
\end{figure*}

The atomic neutral hydrogen (HI) mass represents a fuel reservoir for future star formation, and generally lies in the outskirts of galaxies.  In \simba, the HI content is correlated with SFR even though the simulation assumes that stars form out of H$_2$~\citep{Dave2019}, hence HI provides a bridge between cosmological accretion occurring from the circum-galactic medium and star formation processes in the ISM.
HI content is believed to be governed by relatively recent accretion, and is strongly dependent on environment  \citep[e.g.][]{Mika2015,Kleiner2017,Crain2017,CroneOdekon2018}. 
In the context of galaxy spin acquisition, low mass galaxies were suggested to build their spin in the vorticity-rich  vicinity of filaments, via gas rich infall \citep{laigle2015,Welker2018}. 
One might therefore expect the alignment signal to be stronger for HI rich galaxies.  Here we examine this in \simba.

Figure~\ref{fig:spin_fil_mHI} shows the filament alignment angle PDF for $z=0$ galaxies with high (\textsl{dotted lines}) and low (\textsl{solid lines}) HI content at redshift $z=0$, showing the two extreme stellar mass bins. As we have seen with other quantities, regardless of HI content, low mass galaxies tend to have their spin parallel, and high mass galaxies  perpendicular to their host filament.   

Figure~\ref{fig:residuals_int} (panel c) 
displays the mean cosine of the angle between the spin of galaxies and the direction vectors of their closest filament $\cos \theta_{f-g}$ as a function of HI mass for all galaxies with some content of neutral hydrogen (87\% of \simba galaxies at $z=0$ contain HI).
Galaxies with low HI mass tend to have their spin preferentially perpendicular to filaments' direction, while galaxies with high HI mass are more likely to be aligned with the axis of their host filament. The transition HI mass where the spin flips is at $M_{\rm HI}\approx 10^{9.5} \msun$.

The interesting question then is whether the HI dependence is simply a reflection of the correlation between $M_{\rm HI}$ and \mstar.
To examine this, we also compute the residuals of the cosine of the angle between the spin of galaxies and the direction vectors of their closest filament $\cos \theta_{f-g}$ as a function of HI mass at fixed stellar mass, as shown in Figure~\ref{fig:residuals_int} (panel f).  

Galaxies with high HI content tend to be more aligned with their host filaments compared to average population at same stellar mass, while at low HI mass they are more likely to have their spin  perpendicular. The trend is not markedly different from the panel above, showing that the trend with HI still exists at a fixed \mstar, and hence is not driven by the \mstar alignment dependence.  

Thus it appears that the spin alignment of galaxies is significantly impacted specifically by neutral hydrogen content. Interestingly, despite an overall correlation between HI and SFR in \simba \citep{Dave2019}, the spin alignment dependence on these two properties are markedly different.  This is consistent with the suggestion that the alignment of spin with the local filament is driven by relatively recent accretion spinning up the outskirts of galaxies \citep{pichonetal11}, which may not immediately increase the SFR.

% -------------------------------------------
\section{Impact of local and global environment}
\label{sec:environment}

The results from the previous section, particularly for HI, suggest that anisotropic accretion from the environment plays a specific role in the spin alignment of galaxies and their host filaments (beyond that expected from their host dark halo). In this section we examine the environmental dependence in more detail on small and larger scales, by considering the host halo mass, the density of the nearest filament, and how spin alignment depends on whether a galaxy is a central or a satellite.

% -------------------------------------------
\subsection{Halo mass dependence}
\label{subsec:halo}

We have examined the spin alignment of all halos in Section \ref{sec:haloSpin}. We now focus on the spin alignment of galaxies split by their host halo mass, as a proxy for the depth of the local potential well, to ascertain if it provides an important secondary effect, beyond \mstar, on spin alignments.

\begin{figure}
\centering\includegraphics[width=\columnwidth]{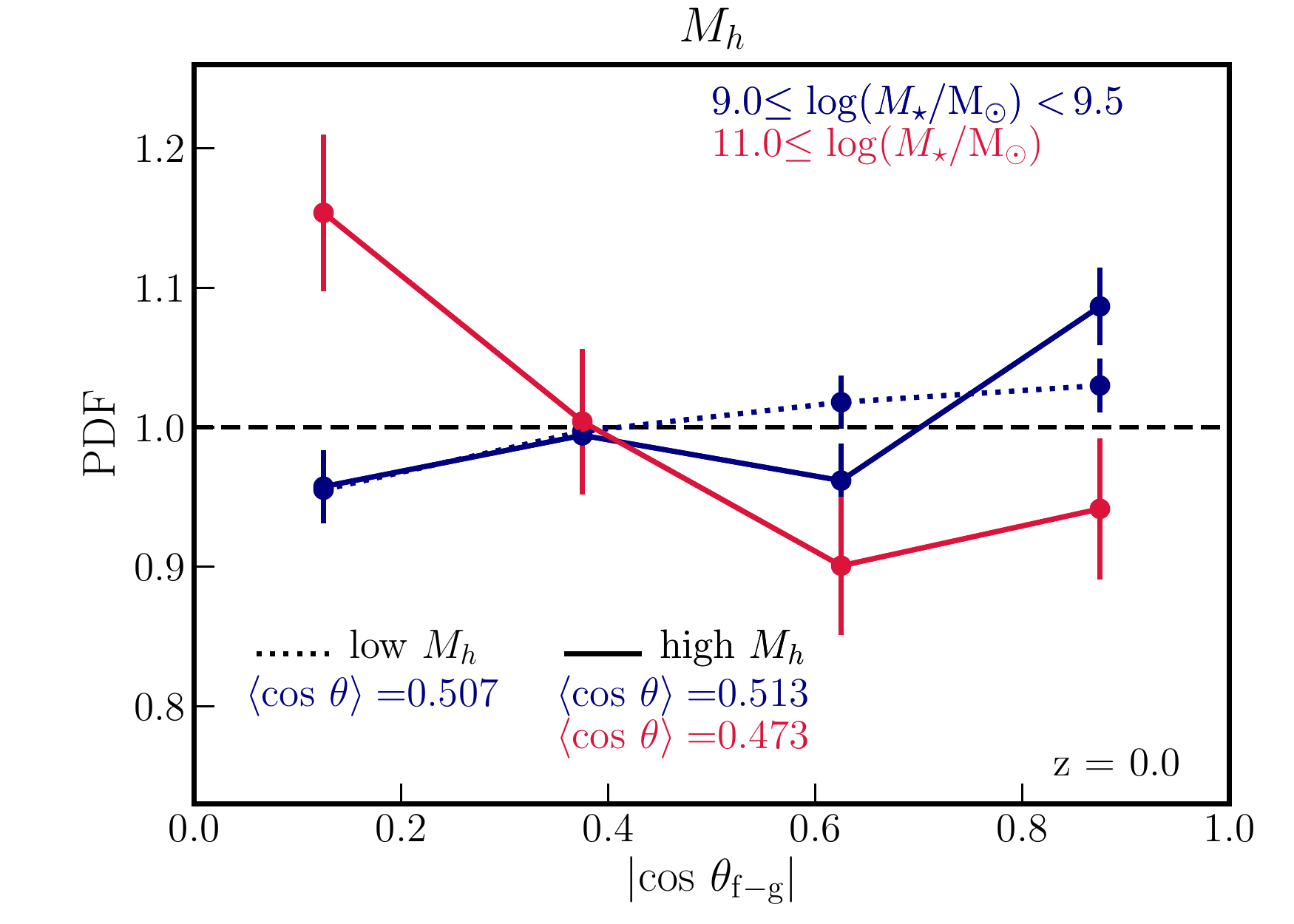}
\caption{Alignment between the spin of galaxies and their closest filaments in different stellar mass bins, as labelled, at redshift $z=0$ living in low (\textsl{dotted lines}) and high (\textsl{solid lines}) mass halos (see Table~\ref{tab:gal_halo} for all \mstar bins). 
The value used to split galaxies corresponds to the median halo mass $10^{11.9}$ \msun. Note that only bins containing more than 100 galaxies are shown.  The error bars represent the Poisson noise. The horizontal black dashed line represents a random distribution. Low mass galaxies tend to align their spin in the direction of their host filament in both low and high mass halos. High mass galaxies are primarily found in massive halos, where their spin is preferentially orthogonal to filaments. The parallel alignment at low stellar mass tend to be dominated by galaxies of high mass halos. 
}
\label{fig:spin_fil_Mh}
\end{figure}

Figure~\ref{fig:spin_fil_Mh} shows the PDF of the cosine of the angle between the spin of galaxies and the direction vectors of their closest filament $\cos \theta_{f-g}$ at redshift $z=0$ for galaxies living in low (\textsl{dotted lines}) and high (\textsl{solid lines}) mass halos.  The value used to split galaxies corresponds to the median halo mass, and we consider only the main halo for each galaxy, not the subhalo.  Because of the underlying stellar-halo mass relation, very few galaxies with mass $\log (\mstar/\msun) \geq 10.5$ live in low mass halos, therefore, the perpendicular alignment with respect to filaments' direction at high stellar mass is driven by galaxies living in massive halos.  However, at low stellar mass, galaxies both of high and low halo mass tend to have their spins aligned with the neighbouring filaments, with a signal stronger for massive halos. 

% ----------------

Figure~\ref{fig:residuals_ext} (panel a) shows
the mean cosine of the angle between the spin of galaxies and their host filaments as a function of host halo mass $M_h$.  Only central galaxies show a clear halo mass dependent flip of the spin, with a transition mass roughly $\log(M_h/\msun) \approx 11.8$. This is in the range of the transition mass for all halos. 
Not surprisingly, due to the lack of a tight correlation between the \mstar of satellites and their host halo mass, there no obvious transition for satellites.

Figure~\ref{fig:residuals_ext} (panel d) shows the residuals of the cosine $\Delta \, \cos \theta_{\rm{f-g}}$ at fixed \mstar, as a function of $M_h$.  This is consistent with zero for all galaxies, suggesting that the alignment signal is driven by \mstar, without any extra variation with the depth of the potential well.

% -------------------------------------------
\subsection{Impact of the host: Central/Satellite}
\label{subsec:cenSat}

We saw in the previous section the satellites do not appear to be aligned with the larger host halo. In this section we separately examine central and satellite spin alignments in various mass bins, in order to investigate the impact of the nature of the host on the measured spin alignments.

\begin{figure}
\centering\includegraphics[width=\columnwidth]{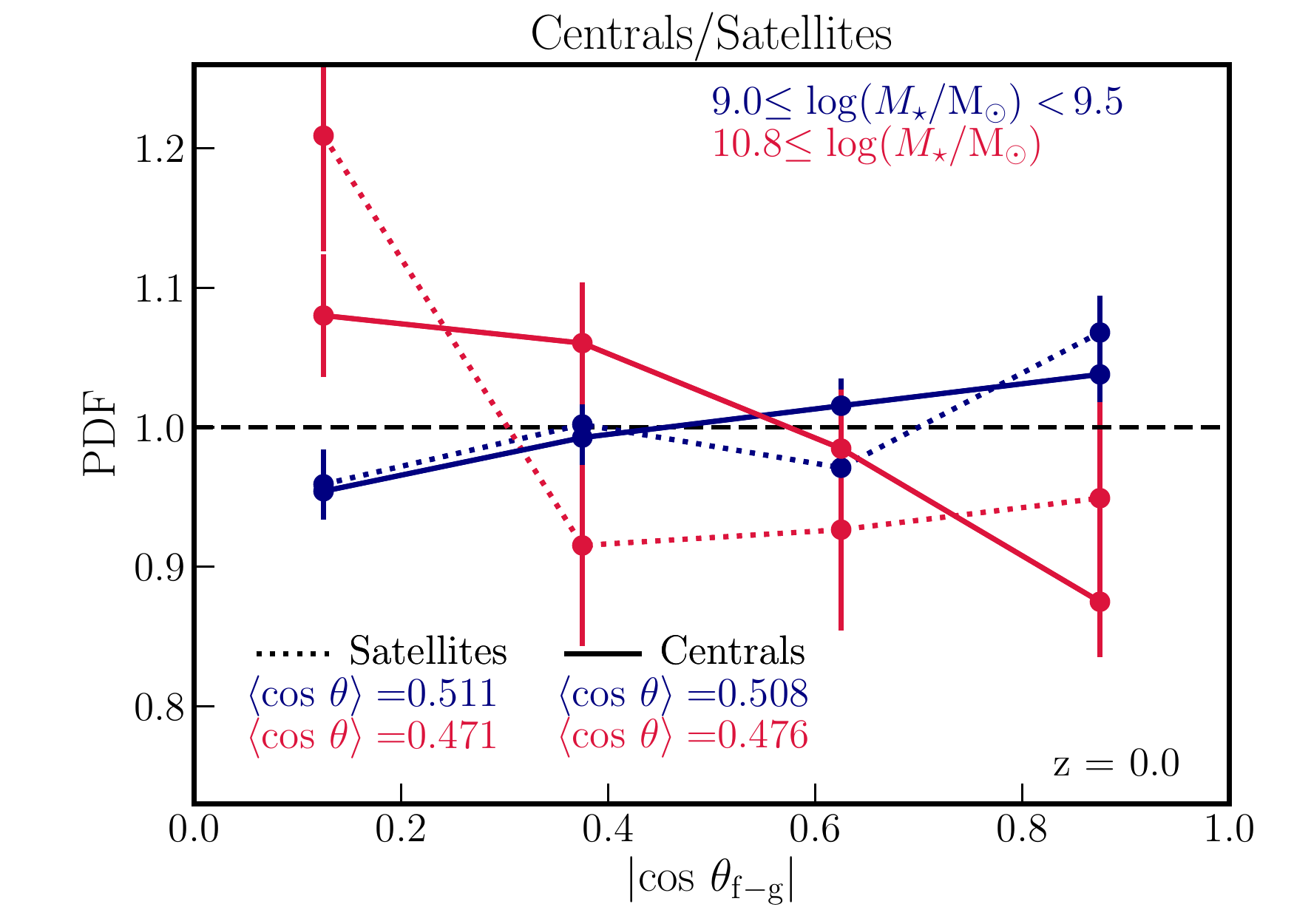}
\caption{Alignment between the spin of galaxies and filaments in different stellar mass bins, as labelled, 
for satellites (\textsl{dotted lines}) and centrals (\textsl{solid lines}) at redshift $z=0$ (see Table~\ref{tab:cen_sat} for all \mstar bins). The error bars represent the Poisson noise. The horizontal black dashed line represents a random distribution. Both central and satellite galaxies show similar alignment signal as the entire galaxy population (note the change of the highest stellar mass bin due to the lack of satellites at such high mass).
However, centrals tend to show slightly stronger alignment signal compared to satellites at the same mass in all but highest stellar mass bins, where the strength is comparable.
}
\label{fig:spin_fil_cenSat}
\end{figure}

Figure~\ref{fig:spin_fil_cenSat} shows the PDF of the cosine of the angle between the spin of galaxies and the direction vectors of their closest filament $\cos \theta_{f-g}$ for centrals and satellites separately at $z=0$. Both galaxy populations show a stellar mass-dependent flip of their spin, with a tendency to be parallel and perpendicular at low and high mass, respectively. The trend for satellites at low \mstar almost exactly mimics that of centrals, while at high masses satellites tend to be more skewed towards perpendicular alignment than centrals. The origin of this transition is discussed in \cite{Welker2018} 
in terms of the kinematics of satellites building-up their spin via quasi-polar flows during their infall into a halo, and subsequently re-orienting their spin through mergers.

% ---------------

% ----------------
\begin{figure*}
\centering\includegraphics[width=\columnwidth]{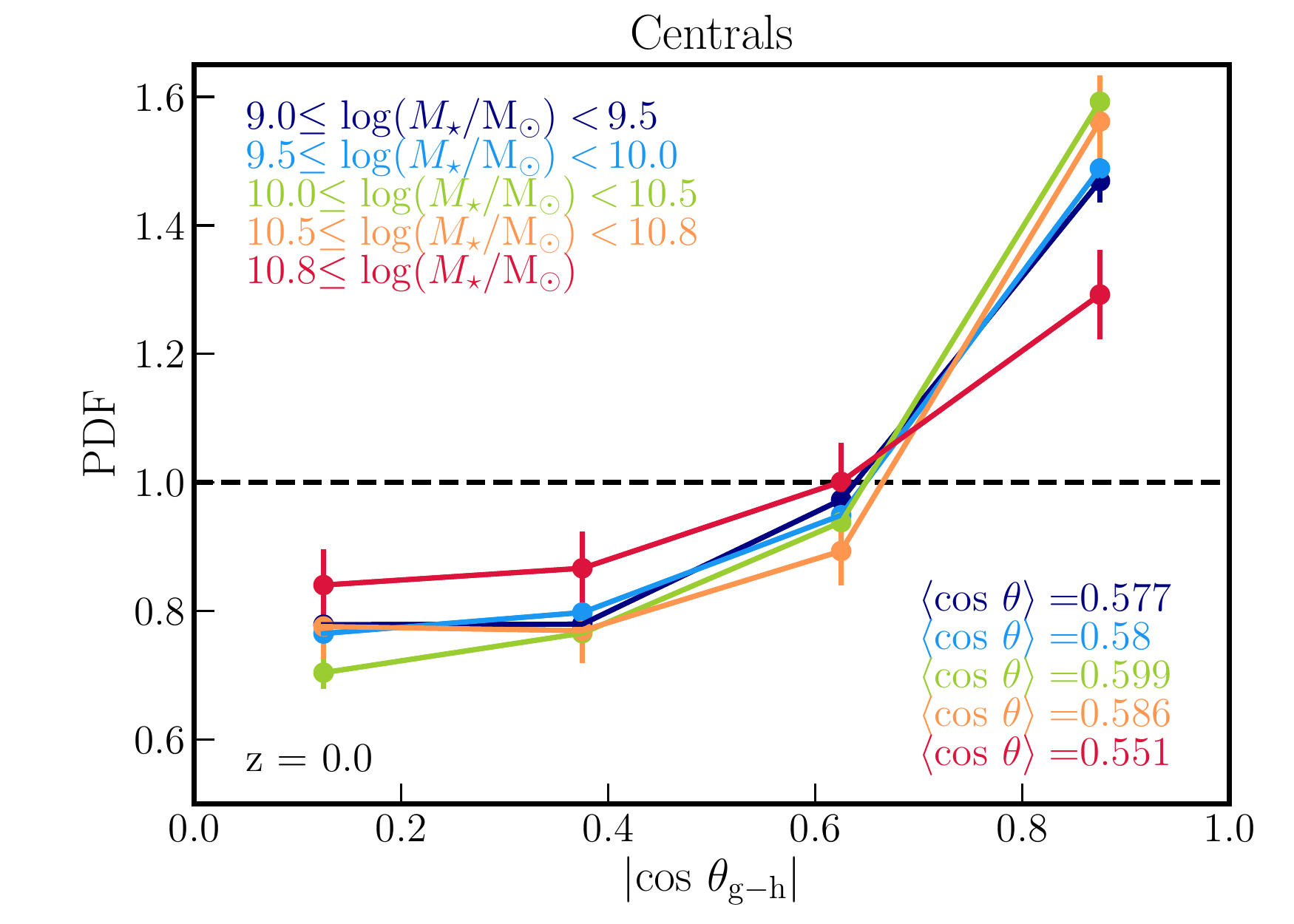}
\centering\includegraphics[width=\columnwidth]{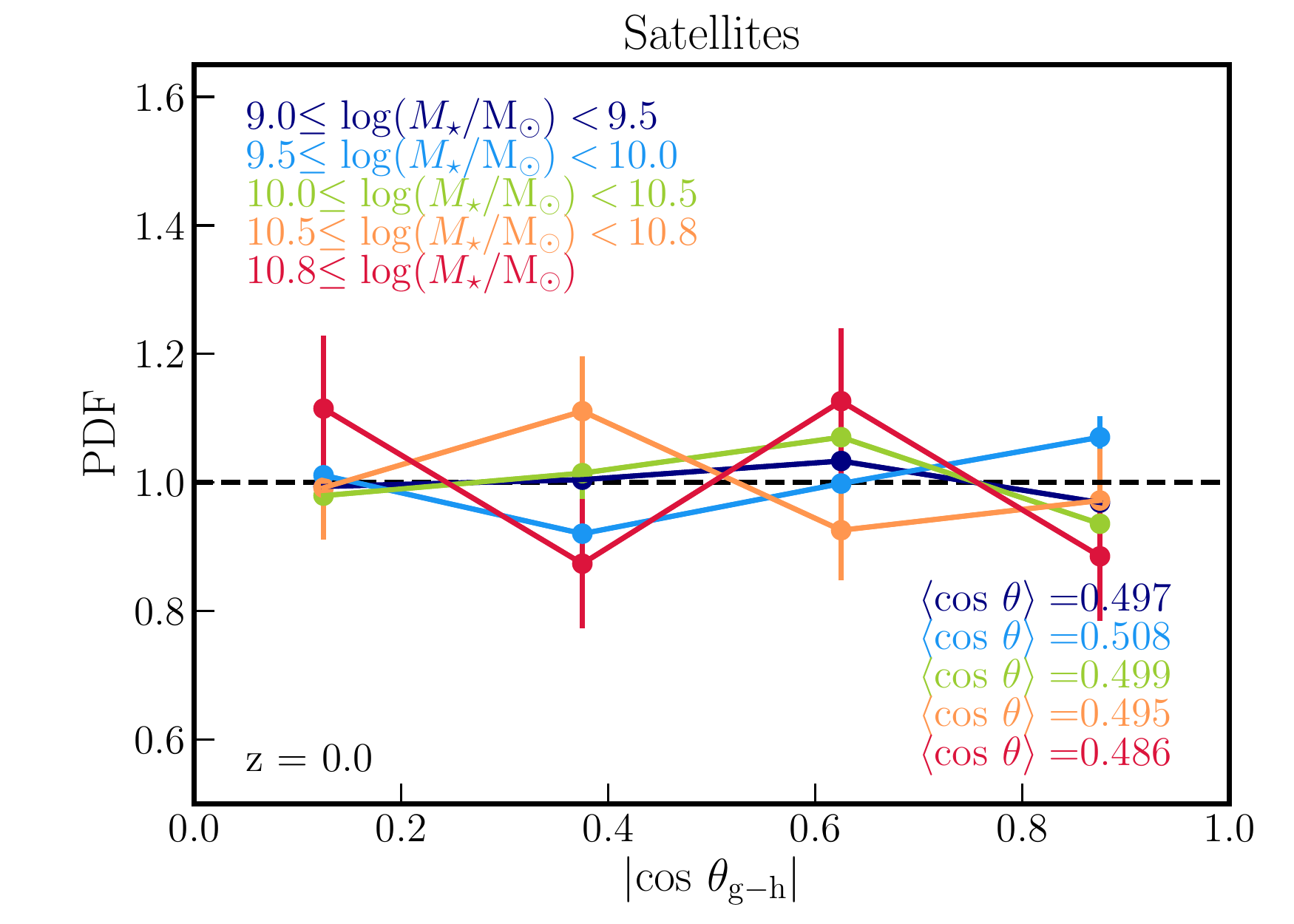}
\caption{Alignment between the spin of galaxies and the spin of their main halo in different stellar mass bins, as labelled, at redshift $z=0$ for centrals (\textsl{left}) and satellites (\textsl{right}). The error bars represent the Poisson noise. The horizontal black dashed line represents a random distribution. The spin of central galaxies is aligned with the spin of their host halo at all masses, while satellites show no correlation.
}
\label{fig:galaxy_halo}
\end{figure*}

% -----------------
\begin{figure*}
\centering\includegraphics[width=\columnwidth]{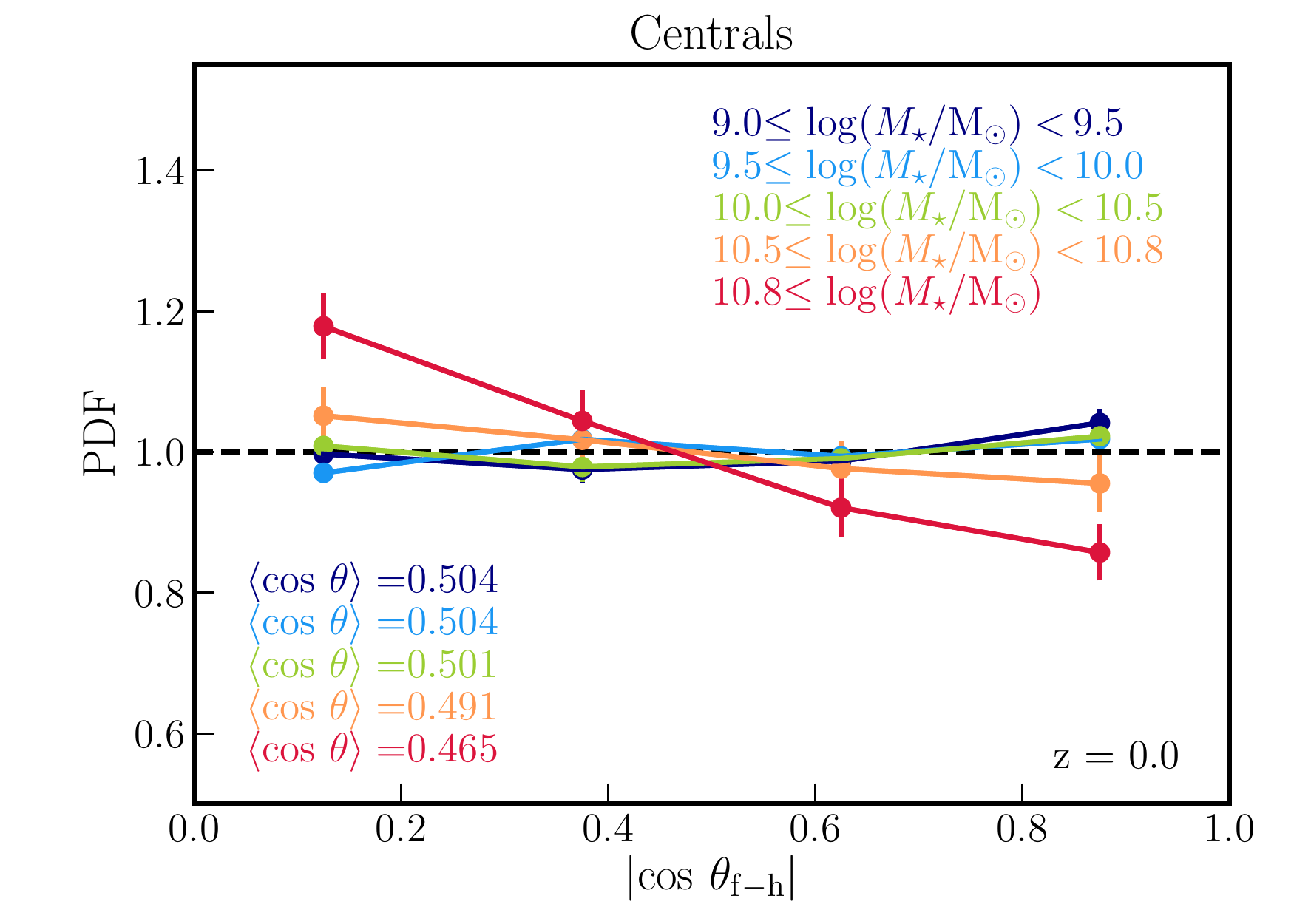}
\centering\includegraphics[width=\columnwidth]{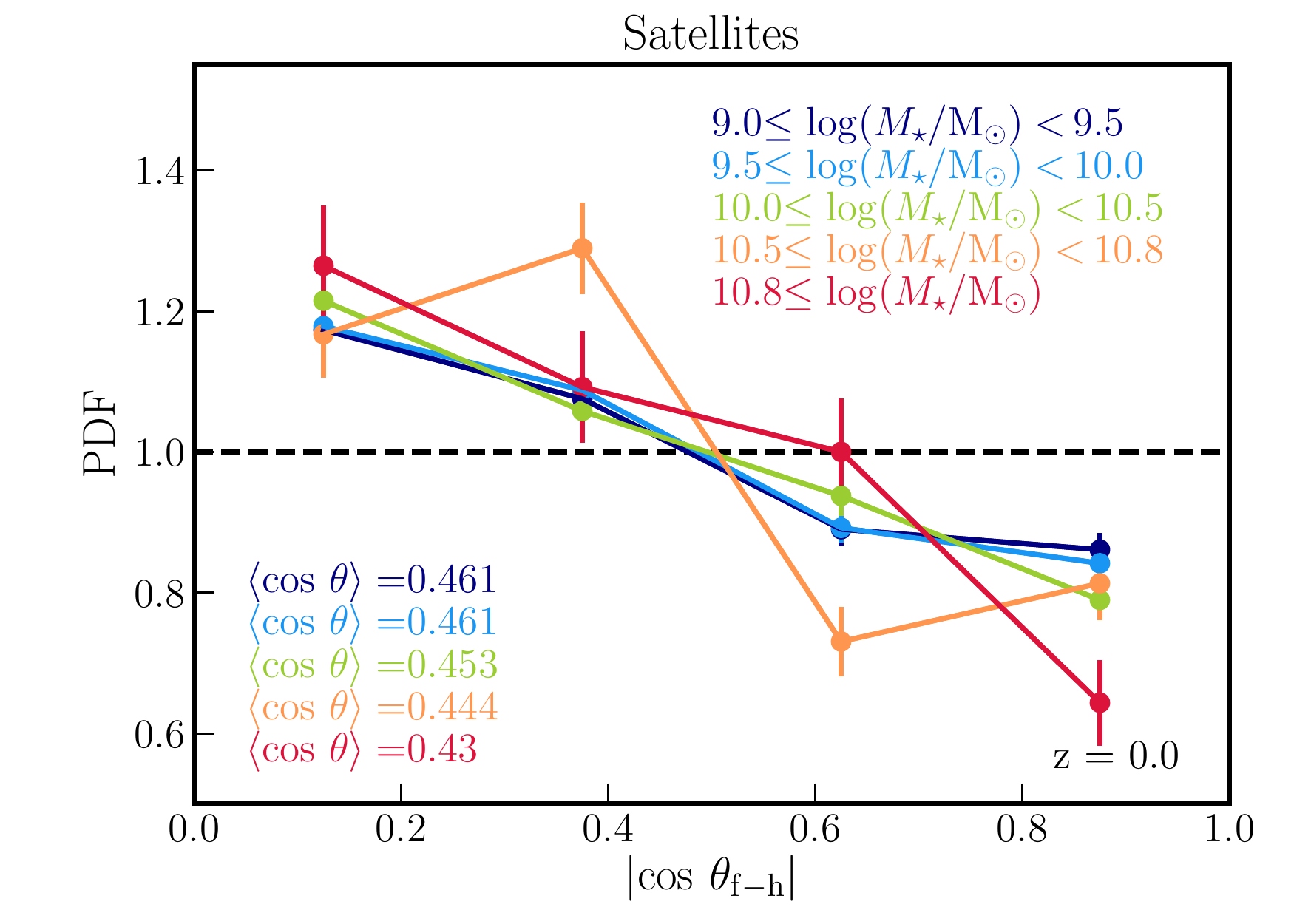}
\caption{Alignment between the spin of halos and galaxies' closest filaments in different stellar mass bins, as labelled, at redshift $z=0$ for centrals (\textsl{left}) and satellites (\textsl{right}).  The error bars represent the Poisson noise. The horizontal black dashed line represents a random distribution. The spin of halos of central galaxies shows a similar mass dependent flip found for galaxies, such that host halos of low mass centrals tend to have their spin aligned with the direction of central's closest filament, while at high mass they are perpendicular. Host halos of satellites have their spin preferentially  perpendicular  to filaments' axes regardless of the mass of satellites. Recall that  only galactic halos are considered here, and their spin orientation is measured with respect to the same filamentary network as for galaxies, i.e. based on the distribution of galaxies.
}
\label{fig:halo_filaments}
\end{figure*}

% -----------------
\begin{figure*}
\centering\includegraphics[width=\columnwidth]{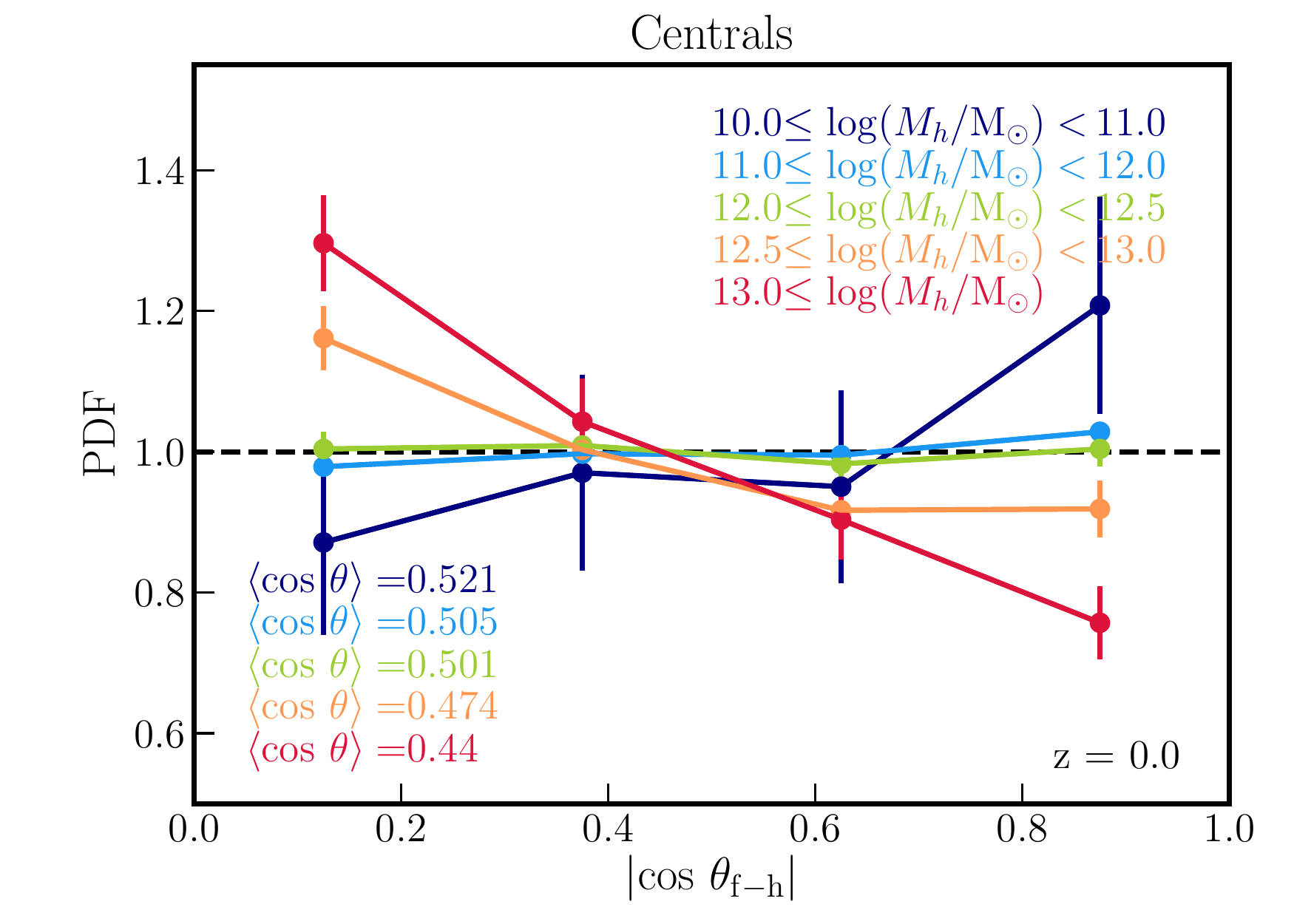}
\centering\includegraphics[width=\columnwidth]{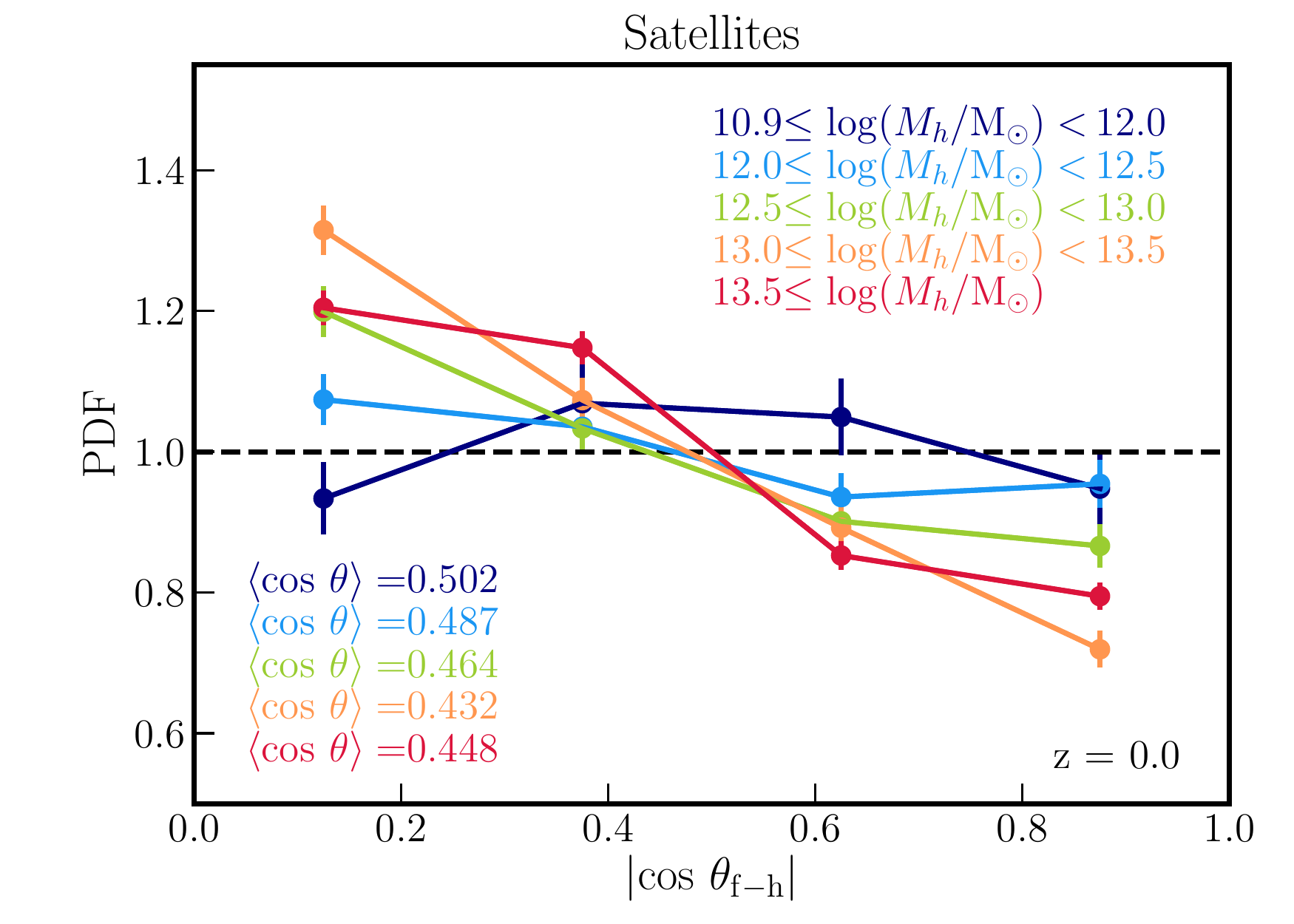}
\caption{As in Figure~\ref{fig:halo_filaments}, but in different halo mass bins, as labelled. Low mass halos of centrals tend to align their spin in the direction of closest galaxy's filament. At high mass, all halos show a clear orthogonal orientation of their spin with respect to filaments. Halos hosting satellites show transition to perpendicular orientation of their spin at lower halo mass compared to halos of centrals ($\log (M_h/\msun) \approx 12.0$ compared to $\log (M_h/\msun) \approx 12.5$).
}
\label{fig:halo_filaments_mHaloBins}
\end{figure*}

We examine this more directly in Figure~\ref{fig:galaxy_halo}, which shows the alignment between the spin of centrals (\textsl{left}) and satellites (\textsl{right}) and the spin of their main halo in different bins of \mstar at $z=0$.  Note that here we consider the absolute value of the cosine of the angle between the spin of galaxies and their main halo, therefore without taking into account the orientation of the spin vectors. Central galaxies have their spin aligned with the spin of their halos at all masses, while the distribution for satellites is consistent with being random in all stellar mass bins.  This is reflected in the mean cosine of the angle being consistent with an alignment of the spin of central galaxies and their host halo ($\left< \cos \theta{\rm g-h}\right>=0.22-0.34$ in the various mass bins), while it shows little correlation between the spin of satellites and that of their parent halo ($\left< \cos \theta{\rm g-h}\right>=0.03-0.05$). 
Hence only the central galaxies' spin are related to their halo's spin, while satellites show very little correlation with their host halo.

Another view of the central/satellite dichotomy is if we consider the alignment between the closest filament and the host halo, as shown in Figure~\ref{fig:halo_filaments} in various \mstar bins. As expected, the centrals follow the host halo trends, with low-mass centrals being weakly parallel aligned, while high-mass ones are most strongly perpendicular.  The interesting trend is for satellites, which shows them strongly perpendicular for all stellar masses.  This partly reflects the halo occupancy distribution, in that the majority of satellites above a given mass live in large halos, which tend to have perpendicular alignment overall.  It is also consistent with
 \cite{aubertetal04} which found that the spin of subhalos lie perpendicular to the halo central separation vector, which typically corresponds
 to the local filament's direction.

A similar trend is found when splitting centrals and satellites by halo mass, as shown in Figure~\ref{fig:halo_filaments_mHaloBins}.  When splitting by halo mass instead of stellar mass, the alignment trend for centrals become much more pronounced.  Interestingly, for satellites, it is still the case that, except for the lowest mass halos, they are perpendicularly aligned.  Indeed, the alignment trends for satellites as a function of halo mass look broadly similar to that for centrals.  Hence the perpendicular alignment of satellites is not purely due to the satellites being predominantly in high-mass halos, but also reflects the satellite spin when it was still a central.

Finally, we can return to Figure~\ref{fig:residuals_int} in order to examine the breakdown of the various second parameter spin alignment trends versus centrals and satellites (\textsl{dotted lines}).  While the trends are broadly similar in alignment as a function of various galaxy properties, they are generally stronger for the central galaxies.  Hence it appears that satellites tend to lose their spin alignment when they fall into another halo, as we saw earlier.

In summary, since satellites tend to be located in denser environments and in more massive halos than centrals at the same stellar mass, the different trends for halos of centrals and satellites seen in Figures~\ref{fig:halo_filaments} and ~\ref{fig:halo_filaments_mHaloBins} could be understood as a consequence of satellites residing near nodes 
where conditional TTT states that their spin should be orthogonal to the filament's direction.

% ------------------------------------------------------
\subsection{Local and filaments' density dependence}
\label{subsec:dens}

Let us now investigate the local and  filaments' density impact on the spin alignment of galaxies. 
In the framework of the TTT, the filament's (and wall's) density is expected to enhance the torque hence the alignment of the spin with respect to the large-scale anisotropic environment.
Conversely, the node's density should also strengthen the perpendicular orientation at high mass.

\begin{figure}
\centering\includegraphics[width=\columnwidth]{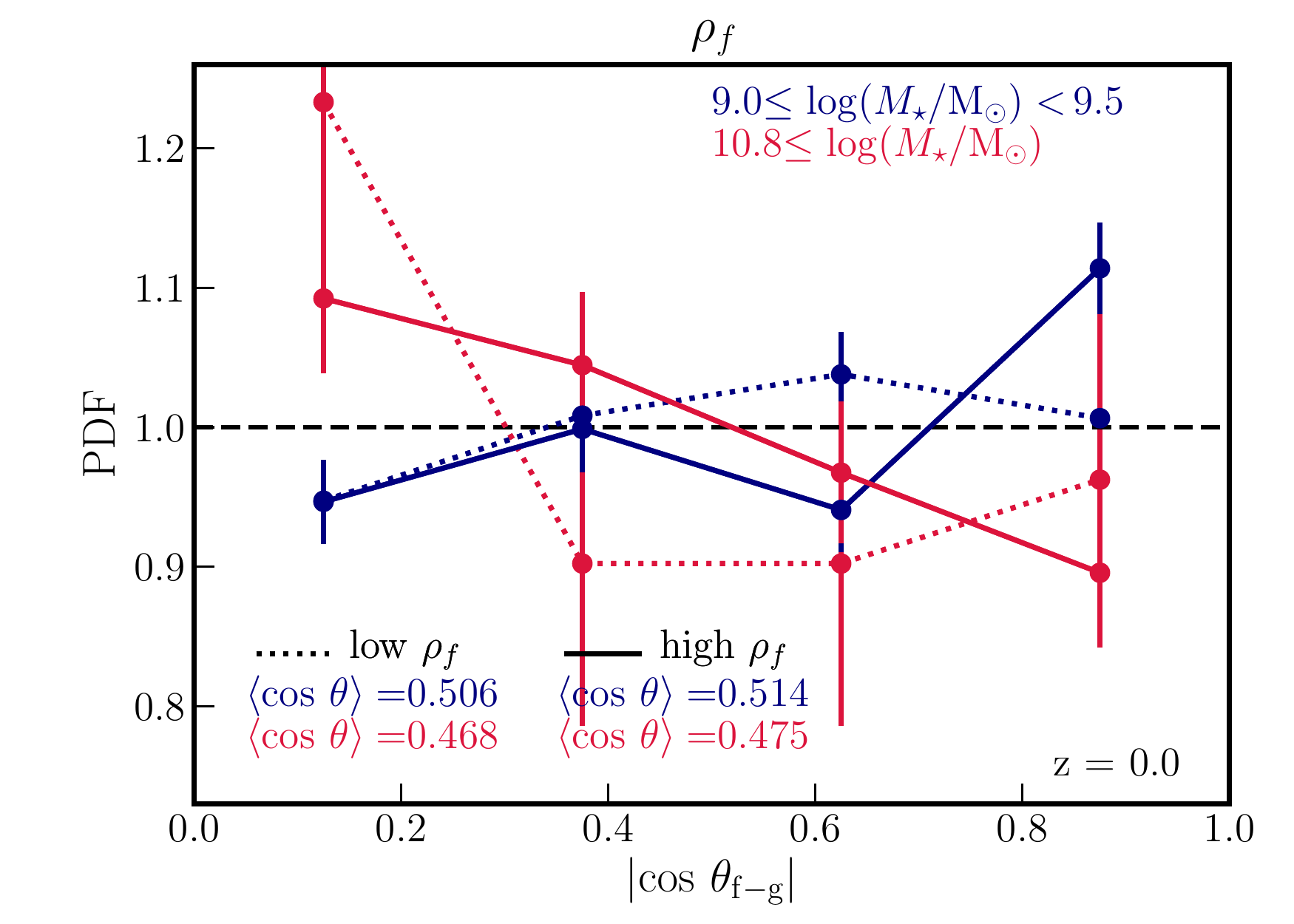}
\caption{Alignment between the spin of galaxies and their closest filaments in the lowest and highest stellar mass bins, as labelled, at redshift $z=0$ for the lowest (\textsl{dotted lines}) and highest (\textsl{solid lines}) filaments' density quartiles, corresponding to $\log (\rho/\rm{Mpc}^{-3}h^{-3}) < -1.1$ and $\log (\rho/\rm{Mpc}^{-3}h^{-3}) > 0.76$, respectively (see Table~\ref{tab:dens} for all \mstar bins). The error bars represent the Poisson noise. The horizontal black dashed line represents a random distribution.
Parallel alignement of the galaxy spin with respect to filaments is at low mass dominated by galaxies associated with high density filaments. The transition from parallel to perpendicular orientation of the spin occurs at higher stellar mass in high density environments. 
}
\label{fig:spin_fil_density}
\end{figure}

Figure~\ref{fig:spin_fil_density} shows the PDF of the cosine of the angle between the spin of galaxies and the direction vectors of their closest filament $\cos \theta_{f-g}$ in the lowest (\textsl{dotted lines}) and the highest (\textsl{solid lines}) filament's density quartile at redshift $z=0$.
As expected, the alignment signal at low \mstar is dominated by galaxies associated with dense filaments. 
Also as expected, at high \mstar, galaxies show statistically significant orthogonal orientation regardless of the filaments' density, since it is the relative node density which now torques it.  Note that the same results are obtained for stellar mass-matched sub-samples of galaxies in low and dense environment, therefore these findings are not driven by low  mass galaxies preferentially occupying low  density large-scale environments (resp. high mass and high density). 

\begin{figure*}
\centering\includegraphics[width=\textwidth]{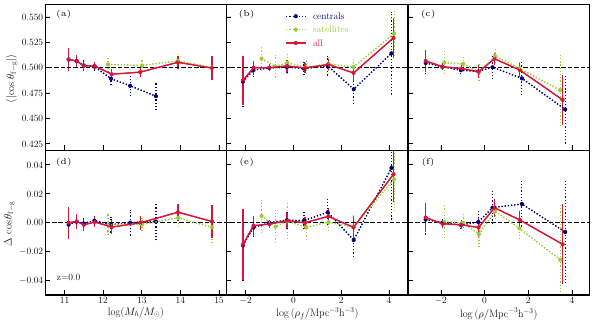}
\caption{ \textsl{Top row:} Mean alignment between the spin of galaxies and filaments as a function of halo mass $M_h$ (\textsl{left}), host filament's density $\rho_f$ (\textsl{middle}) and local density $\rho$ (\textsl{right}) of galaxies at $z=0$. 
Galaxies living in low mass halos tend to align their spin with the filament's axis, at intermediate halo mass they tend to have their spin in perpendicular direction, while at high halo mass only centrals continue to spin in perpendicular direction to the filaments. 
Galaxies associated with low density filaments have their spin preferentially in perpendicular direction to filaments, while at high densities, there is a clear trend for galaxies to align their spin with their host filament. 
Galaxies living in high density regions tend to have their spin in perpendicular direction to the filament's axis.
\textsl{Bottom row:} Residuals of the cosine of the angle between the spin of galaxies and the direction vectors of their closest filament $\cos \theta_{f-g}$ at fixed \mstar. 
Trends seen as a function of halo mass are entirely driven by \mstar, while 
parallel spin-filament orientation of galaxies is clearly dominated by galaxies associated with high density filaments. Perpendicular spin-filament orientation of galaxies is marginally found to be dominated by galaxies living in highest density regions.
}
\label{fig:residuals_ext}
\end{figure*}

This is consistent with the filament's density dependence of the alignments displayed on Figure~\ref{fig:residuals_ext} (panel b), showing that the spin of galaxies associated with high density filaments tend to be aligned with their axis, while galaxies tend to have their spin perpendicular with respect to low density filaments. This trend is not driven by stellar mass alone, as residuals at fixed \mstar (Figure~\ref{fig:residuals_ext}, panel e) show similar trends suggesting in particular that the parallel spin-filament alignment signal is  also driven by high density filaments, imposing stronger tides.  

Finally, we examine the spin alignment dependence on the local density at the galaxy's position, rather than the density of the nearest filament. Interestingly, Figure~\ref{fig:residuals_ext} (panel c) shows that galaxies in high local density regions tend to have their spin preferentially  perpendicular  to their host filament, while in low local density regions, they tend to have their spin  parallel.
Residuals at fixed \mstar are close to zero (Figure~\ref{fig:residuals_ext}, panel f), suggesting that  stellar mass is driving the spin alignments, regardless of the local density.

The impact of the local density can be explained by differences in the position of these galaxies with respect to the cosmic web. 
Because of density gradients along filaments toward nodes, galaxies of the same filament that are further away typically have lower local density, compared to galaxies located in the vicinity of the nodes. 
Indeed, when considering the local density, the effect at high density is enhanced for galaxies in low \mstar bin, while high mass galaxies are clearly perpendicular orientation, even at the highest densities. 
Thus the positive residuals at high filament's density are driven by galaxies further away from nodes. This is consistent with the interpretation that mergers (or equivalently accretion along the filaments), frequent in the nodes of the cosmic web, are driving the spin flips. We conclude that  the exact 3D position of galaxies in the frame of the cosmic web is important to interpret the observed trends, as galaxies and their properties trace the geometry of the bulk flow within that frame \citep[see e.g.][]{Kraljic2019}.

% ---------------

% -------------

% -----------------------------------------------
\section{Summary and Discussion}
\label{sec:summary}

We have investigated the correlation between the spin orientation of galaxies and halos and their large-scale anisotropic environment using the state-of-the-art cosmological hydrodynamical simulation \simba. The orientation of the angular momentum was measured relative to the direction of filaments and walls  identified with the topological extractor \disperse.  Our main focus was on its evolution as a function  of stellar mass and redshift, as well as its dependence on secondary internal parameters such as morphology, HI content, star formation activity, and external parameters such as their halo mass, filament's density and the central/satellite dichotomy.
Our principal findings are as follows.
\begin{itemize}
\item{\textsl{Halos:}} Halos show a strong alignment of their spin parallel to the filaments and walls at low masses and orthogonal at high masses, with a transition occurring at around $\log (M_h/\msun) \approx 12.0 \pm 0.5$ at $z=0$ for filaments, and at slightly higher mass for walls. This transition mass decreases with increasing redshift. 
\item{\textsl{Galaxies:}} Galaxies' spin flip occurs at a corresponding mass of $\log (\mstar/\msun) \approx 10 \pm 0.5$.
\item{\textsl{Morphology:}} 
The perpendicular orientation is driven by high-mass dispersion-dominated systems, while rotation-dominated galaxies drive the parallel alignment at low masses. There is an additional weak trend due to morphology, beyond the trend established by \mstar alone.
\item{\textsl{Star formation:}} 
Star-forming galaxies are preferentially parallel to filaments and quiescent galaxies tend to orient themselves perpendicularly, however this trend can be explained purely by its \mstar dependence.
\item{\textsl{HI mass:}} Interestingly, spin alignment is found to correlate significantly with HI content. Even at a given stellar mass, galaxies with high HI mass tend to align their spins with the axis of the filament while the spin of low HI-mass galaxies is more likely to be perpendicular to the direction of the closest filament.  This suggests that recent accretion drives up both galaxy spin as well as HI content, and that the accretion appears to have an angular momentum that is parallel to the filament.
\item{\textsl{Host halo mass:}} Low mass galaxies align their spin to filaments in both low and high mass (main) halos. High mass galaxies are all found in  high mass halos, therefore they all display a preferential orthogonal orientation of their spin with respect to filaments. 
However, no additional dependence of the spin-filament orientation on the halo mass is detected beyond the \mstar dependence.
\item{\textsl{Centrals/Satellites:}} Central and satellite galaxies both show a stellar mass dependent flip of their spin with respect to filaments at $z=0$. 
Due to the tight (main) halo mass--stellar mass correlation for centrals, this population of galaxies shows also a halo mass dependent flip of their spin, with centrals living in low (high) mass halos aligning their spin in parallel (perpendicular) direction to the filaments' axes. Spin-filament alignment for satellites does not show any correlation with the mass of their main halo. At fixed \mstar, no residuals in the alignment signal as a function of halo mass are detected, neither for centrals nor for satellites.
\item{\textsl{Centrals/Satellites vs. halo spin:}} Central galaxies tend to align their spin with the spin of their host halos at all stellar masses, while satellites show no correlation. Halos of centrals align their spin with the filaments at low stellar and halo mass, and have their spin perpendicular at high stellar and halo mass. Halos of satellites  have their spin clearly perpendicular independently of stellar mass, and in all but the lowest halo mass bin.
\item{\textsl{Local and nearby filaments' density:}}  
The alignment signal at low \mstar is dominated by galaxies of high density filaments, while at high \mstar, spin of galaxies is orthogonal to filaments at all densities.   
As a result, the residuals at fixed \mstar
suggest that at low filaments' densities, galaxy spins tend to be orthogonal, and at high densities parallel to the host filaments. When considering instead the density at the location of galaxies, residuals at fixed \mstar are close to zero, suggesting that  stellar mass is enough to account for the observed spin flip.
\end{itemize} 
  
The alignment of the spin of halos with respect to filaments of the cosmic web has received a lot of attention in the past, in part to test predictions from tidal torque theory, and the canonical assumption used in e.g. semi-analytic models that the spin of the galaxy follows the direction of its host halo. In contrast, studies of the alignment of the {\it galaxy} spin in the context of large scale structure have emerged only recently, motivated for example because intrinsic alignments are a source of contamination for weak lensing-based dark energy surveys \citep{Chisari2017}.

Our results for halos' spin showing a flip in spin orientation from low to high masses, and the corresponding transition mass, are in a good agreement with trends seen in both dark matter only simulations \citep[e.g.][]{AragonCalvo2007,Codis2012,GaneshaiahVeena2018}, and most hydrodynamical simulations \citep[e.g.][]{Codis2018,GaneshaiahVeena2019}.
Analogously, the stellar mass dependent flip of galactic spin found in the present work is also consistent with findings of \cite{Welker2014,Codis2018} and \cite{Wang2018}, at $z \leq 2$ and $z=0$, respectively. While we did not test the specific role of mergers as in \cite{Welker2014}, their results that mergers do not play a role in low-mass spin alignments is consistent with the importance of accretion as traced by HI content for driving the spin alignment in \simba.
The lack of detection of a clear transition reported by \cite{GaneshaiahVeena2019} was argued to be a consequence of the properties of the underlying filament, with galaxies in thinner filaments having their spins more likely perpendicular to the filament's axis, compared to galaxies of similar mass in thicker filaments.
A straightforward interpretation of this result is the multi-scale nature of the problem. At fixed halo mass, changing the thickness of filaments is equivalent to changing the smoothing scale defining the filament, hence changing the mass of non-linearity against which that mass must be compared\footnote{The denser the filament the thicker and therefore the larger the transition mass \citep[see Fig 17 of][]{Codis2015} as it corresponds to the mass enclosed in the sphere of radius one-half 
of the radius of the filament.  At fixed halo mass, the lighter the filament the more perpendicular the alignment. Alternatively and equivalently, if the halo is smaller than the quadrant 
of vorticity defined by its thickness, it will have its spin aligned with the filament \citep{laigle2015}.}. 
Conditional TTT predicts that if it is below the corresponding mass of non-linearity (thick filament), the spin tends to be parallel, and if it is above, the spin is perpendicular. 
To confirm this, since filament thickness is not something that is topologically defined, hence not characterized using \disperse, we used the filament density as a proxy instead. With this proxy, our findings are consistent with conditional TTT predicting stronger impact of large scale tides on the galaxy spin orientation in denser filaments.

Regarding the dependence of the spin-filament alignment on the internal properties of galaxies, our finding that the parallel alignment tend to be driven by galaxies with high \vsig (rotation dominated galaxies) while the perpendicular alignment signal is dominated by low \vsig population (with elliptical morphologies) is in agreement with results of \cite{Codis2018}. The dependence on star formation activity can be explained purely as a stellar mass effect, in qualitative agreement with \cite{Wang2018}, when splitting galaxies into blue and red populations based on their $g-r$ color.  The stronger signal for morphology versus star formation activity suggests that spin is more directly related to the former.

Consistently with previous studies \citep[e.g.][]{Codis2018}, the spin of satellites in \simba is found to be uncorrelated with the spin of their main halo, while the spin of centrals is much better correlated with that of their halos. It was suggested by these authors that this may be an indication that satellites lose the memory of the filaments from which they emerged during virialisation. 
That said, we do find in \simba that satellites still show a stellar mass dependent flip, in tension with the findings of \cite{Codis2018} at low redshifts, showing no transition and no mass dependence.  This may reflect a difference between these simulations in how long satellites retain memory of their original halo's spin, i.e. how much merging and harassment they undergo as satellites.

Interestingly, we find that the parent halos of satellites have their spin clearly perpendicular to the filaments' direction, independently of their stellar mass. 
Given that satellites tend to live in more massive halos than the centrals at the same stellar mass, this could be a signature of the merger induced perpendicular orientation of halo spins at higher mass.
Beyond the processes satellites undergo as they interact with their hosts and other satellites, they are also more likely to be influenced by strong AGN feedback from their massive central, potentially modifying their spin orientation. The implementation of the AGN feedback in the \simba simulation differs significantly from the prescription used in other simulations (e.g. \ill, \eagle or \hagn), as \simba always uses kinetic bipolar outflows for all black hole feedback. Conversely, the more spherical (thermal) feedback implemented in simulations such as \eagle and for moderate-sized black holes in \ill and \hagn may more efficiently destroy the cosmic flows feeding satellites with angular momentum-rich cold gas, and building their own spin parallel to their embedding filaments \citep{duboisetal12}.

There has also been some controversy regarding hydrodynamics methodology. As it happens, the results of previous studies on  spin alignments in cosmological hydrodynamical simulations seem to fall in two categories depending on the implemented numerical technique. 
Works using smoothed particle hydrodynamics simulations typically did not detect the mass dependent flip of the spin, in contrast to those analysing simulations using adaptive mesh refinement codes. 
\simba employs an ALE-like (Arbitrary Lagrangian Eulerian) code for hydrodynamics, which is fully adaptive in a Lagrangian sense but uses a Riemann solver rather than smoothed pressures to compute forces, and does not include an artificial viscosity like SPH codes.  Our results are more consistent with findings relying on AMR codes using the same cosmic web finder, indicating that (i) the details in modelling of hydrodynamics may play an important role in preserving the subtle interplay between the larger scale cosmic web and the internal dynamics of galaxies; and (ii) AMR grid locking cannot be the sole source of spin alignment at low mass.  In addition, different algorithms used to identify the cosmic web may impact the quantitative signal of spin alignments with large-scale structure.  Future comparisons to observations should strive to employ similar techniques, such as applying \disperse to galaxy redshift surveys.

Finally, our results are globally consistent with conditional tidal torque theory constrained to the vicinity of filaments and walls \citep{Codis2015a}.
As this theory strictly applies to dark matter halos only, 
finding qualitatively similar results for galaxies suggests that in spite of a variety of baryonic processes not accounted for in this theoretical framework, galaxies in fact continue to be impacted by the dynamics of the large-scale cosmic flows from which their halos originated. 
Our results particularly highlight the correlation of spin alignment with respect to the HI content as further evidence for the role of the cosmic flows in feeding angular momentum-rich gas to young galaxies \citep{pichonetal11,stewart2011}.  This also suggests that HI could play an important role in identifying the expected orientation of galaxies in surveys where galaxy shapes are poorly resolved, such as in some upcoming weak lensing surveys with Large Synoptic Survey Telescope (LSST\footnote{https://www.lsst.org/}).

% ---------------------------------------------

\section*{Acknowledgements}
\simba was run on the DiRAC@Durham facility managed by the Institute for Computational Cosmology on behalf of the STFC DiRAC HPC Facility. The equipment was funded by BEIS capital funding via STFC capital grants ST/P002293/1, ST/R002371/1 and ST/S002502/1, Durham University and STFC operations grant ST/R000832/1. DiRAC is part of the National e-Infrastructure.
This work is partially supported by the Spin(e) grants ANR- 13-BS05-0005 (http://cosmicorigin.org) of the French Agence Nationale de la Recherche. We thank S. Rouberol for smoothly running the {\sc Horizon} cluster,
hosted by the Institut d'Astrophysique de Paris, where some 
of the postprocessing was carried out, and T. Sousbie for \disperse. KK thanks Elisa Chisari, Sandrine Codis and Clotilde Laigle for fruitful comments and discussions.
RD acknowledges support from the Wolfson Research Merit Award program of the U.K. Royal Society. 

%%%%%%%%%%%%%%%%%%%%%%%%%%%%%%%%%%%%%%%%%%%%%%%%%%



\bibliographystyle{mnras}
\bibliography{author} % if your bibtex file is called example.bib

%%%%%%%%%%%%%%%%%%%%%%%%%%%%%%%%%%%%%%%%%%%%%%%%%%

%%%%%%%%%%%%%%%%% APPENDICES %%%%%%%%%%%%%%%%%%%%%
% ======
\appendix

% ------------------------------------------
\section{KS test probabilities}
\label{app:KS}

This Appendix provides the measure of  
the statistical significance in terms of Kolmogorov-Smirnov test for Figures presented in the main text. 
Tabels~\ref{tab:mass}-\ref{tab:halo_gal_fil_haloBins} contain information about the number of galaxies/halos, mean cosine of angle between the spin of galaxies/halos and their host filaments/walls and the KS probability $p_{\rm KS}$ that the sample is drawn from a uniform distribution. %, for Figures~\ref{fig:spin_fil}-\ref{fig:halo_filaments_mHaloBins}.

% ------------------
\begin{table*}
\centering
\begin{threeparttable}
\caption{Redshift $z$, number of halos $N_{\rm gal}$, average $\cos \theta$ and the KS probability $p_{\rm KS}$ that the sample is drawn from a uniform distribution for Figure~\ref{fig:spin_halo_fil_halo}}
\label{tab:halos}
\begin{tabular*}{0.7\textwidth}{@{\extracolsep{\fill}}lccccc}
\hline
\hline
& $z$ &\mstar & $N_{\rm gal}$ & $\left < \cos \theta \right>$ & $p_{\rm KS}$ \\
\hline
\hline
\multirow{15}{*}{Filaments} & \multirow{5}{*}{0.0} & 10.0 $\leq \log (M_h/\msun) < $ 10.5 & 376924 & 0.508 & 2.86$\times 10^{-48}$ \\
& & 10.5 $\leq \log (M_h/\msun) < $ 11.0 & 148862  & 0.508 &  1.0$\times 10^{-24}$\\
& & 11.0 $\leq \log (M_h/\msun) < $ 11.5 & 53036 & 0.508 & 7.5$\times 10^{-7}$ \\
& & 11.5 $\leq \log (M_h/\msun) < $ 12.0 & 19470 & 0.505 & 0.02\\
& & $12.0 \leq \log (M_h/\msun) < $ 12.5 & 6520 & 0.496 & 0.09 \\
& & $12.5 \leq \log (M_h/\msun)$ & 3410 & 0.486 & 0.01 \\
\cline{2-6}
& \multirow{5}{*}{1.0} & 10.0 $\leq \log (M_h/\msun) < $ 10.5 & 442274 & 0.508 & 9.8$\times 10^{-62}$ \\
& & 10.5 $\leq \log (M_h/\msun) < $ 11.0 & 173722  & 0.507 &  6.3$\times 10^{-22}$\\
& & 11.0 $\leq \log (M_h/\msun) < $ 11.5 & 56670 & 0.506 & 3.6$\times 10^{-5}$ \\
& & 11.5 $\leq \log (M_h/\msun) < $ 12.0 & 20332 & 0.502 & 0.32\\
& & $12.0 \leq \log (M_h/\msun) < $ 12.5 & 7090 & 0.487 & 2.7$\times 10^{-5}$ \\
& & $12.5 \leq \log (M_h/\msun)$ & 2510 & 0.473 & 6.4$\times 10^{-6}$ \\
\cline{2-6}
& \multirow{5}{*}{2.0} & 10.0 $\leq \log (M_h/\msun) < $ 10.5 & 467958 & 0.508 & 9.3$\times 10^{-61}$ \\
& & 10.5 $\leq \log (M_h/\msun) < $ 11.0 & 172296  & 0.507 &  3.5$\times 10^{-19}$\\
& & 11.0 $\leq \log (M_h/\msun) < $ 11.5 & 52518 & 0.502 & 4.5$\times 10^{-3}$ \\
& & 11.5 $\leq \log (M_h/\msun) < $ 12.0 & 16178 & 0.496 & 0.13\\
& & $12.0 \leq \log (M_h/\msun) < $ 12.5 & 4792 & 0.477 & 2.7$\times 10^{-7}$ \\
& & $12.5 \leq \log (M_h/\msun)$ & 1150 & 0.459 & 5.2$\times 10^{-6}$ \\
\hline
\multirow{15}{*}{Walls} & \multirow{5}{*}{0.0} & 10.0 $\leq \log (M_h/\msun) < $ 10.5 & 376924 & 0.491 & 1.1$\times 10^{-71}$ \\
& & 10.5 $\leq \log (M_h/\msun) < $ 11.0 & 148862  & 0.49 &  1.1$\times 10^{-31}$\\
& & 11.0 $\leq \log (M_h/\msun) < $ 11.5 & 53036 & 0.494 & 3.1$\times 10^{-8}$ \\
& & 11.5 $\leq \log (M_h/\msun) < $ 12.0 & 19470 & 0.502 & 0.14\\
& & $12.0 \leq \log (M_h/\msun) < $ 12.5 & 6520 & 0.508 & 0.06 \\
& & $12.5 \leq \log (M_h/\msun)$ & 3410 & 0.506 & 0.19 \\
\cline{2-6}
& \multirow{5}{*}{1.0} & 10.0 $\leq \log (M_h/\msun) < $ 10.5 & 442274 & 0.489 & 2.6$\times 10^{-124}$ \\
& & 10.5 $\leq \log (M_h/\msun) < $ 11.0 & 173722  & 0.489 &  2.9$\times 10^{-49}$\\
& & 11.0 $\leq \log (M_h/\msun) < $ 11.5 & 56670 & 0.493 & 3.8$\times 10^{-11}$ \\
& & 11.5 $\leq \log (M_h/\msun) < $ 12.0 & 20332 & 0.5 & 0.14\\
& & $12.0 \leq \log (M_h/\msun) < $ 12.5 & 7090 & 0.508 & 0.07 \\
& & $12.5 \leq \log (M_h/\msun)$ & 2510 & 0.508 & 0.05 \\
\cline{2-6}
& \multirow{5}{*}{2.0} & 10.0 $\leq \log (M_h/\msun) < $ 10.5 & 467958 & 0.487 & 8.4$\times 10^{-194}$ \\
& & 10.5 $\leq \log (M_h/\msun) < $ 11.0 & 172296  & 0.489 & 8.4$\times 10^{-13}$\\
& & 11.0 $\leq \log (M_h/\msun) < $ 11.5 & 52518 & 0.493 & 0.02 \\
& & 11.5 $\leq \log (M_h/\msun) < $ 12.0 & 16178 & 0.497 & 0.13\\
& & $12.0 \leq \log (M_h/\msun) < $ 12.5 & 4792 & 0.508 & 0.03 \\
& & $12.5 \leq \log (M_h/\msun)$ & 1150 & 0.516 & 0.02 \\
\hline
\hline
\end{tabular*}
\end{threeparttable}
\end{table*}

% ------------------
\begin{table*}
\centering
\begin{threeparttable}
\caption{Redshift $z$, number of galaxies $N_{\rm gal}$, average $\cos \theta$ and the KS probability $p_{\rm KS}$ that the sample is drawn from a uniform distribution for Figure~\ref{fig:spin_fil}}
\label{tab:mass}
\begin{tabular*}{0.7\textwidth}{@{\extracolsep{\fill}}lccccc}
\hline
\hline
& $z$ &\mstar & $N_{\rm gal}$ & $\left < \cos \theta \right>$ & $p_{\rm KS}$ \\
\hline
\hline
\multirow{15}{*}{Filaments} & \multirow{5}{*}{0.0} & 9.0 $\leq \log (\mstar/\msun) < $ 9.5 & 16566 & 0.509 & 1.7 $\times 10^{-3}$ \\
& & 9.5 $\leq \log (\mstar/\msun) < $ 10.0 &  26996 & 0.501 & 0.7 \\
& & 10.0 $\leq \log (\mstar/\msun) < $ 10.5 & 13672 & 0.499 & 0.037 \\
& & 10.5 $\leq \log (\mstar/\msun) < $ 11.0 & 5168 & 0.482 & 9.4 $\times 10^{-5}$\\
& & $11.0 \leq \log (\mstar/\msun)$ & 1470 & 0.473 & 2.9 $\times 10^{-4}$ \\
\cline{2-6}
& \multirow{5}{*}{1.0} & 9.0 $\leq \log (\mstar/\msun) < $ 9.5 & 10734 & 0.503 & 0.15\\
& & 9.5 $\leq \log (\mstar/\msun) < $ 10.0 & 11230 & 0.502 & 0.85\\
& & 10.0 $\leq \log (\mstar/\msun) < $ 10.5 & 8946 & 0.494 & 0.089\\
& & 10.5 $\leq \log (\mstar/\msun) < $ 11.0 & 4498 & 0.489 & 0.063\\
& & $11.0 \leq \log (\mstar/\msun)$ & 1258 & 0.47 & 1.6$\times 10^{-4}$\\
\cline{2-6}
& \multirow{5}{*}{2.0} & 9.0 $\leq \log (\mstar/\msun) < $ 9.5 & 7778 & 0.506 & 0.014\\
& & 9.5 $\leq \log (\mstar/\msun) < $ 10.0 & 5890 & 0.499 & 0.25\\
& & 10.0 $\leq \log (\mstar/\msun) < $ 10.5 & 3188 & 0.497 & 0.56\\
& & 10.5 $\leq \log (\mstar/\msun) < $ 11.0 & 1670 & 0.494 & 0.24\\
& & $11.0 \leq \log (\mstar/\msun)$ & 682 & 0.466 & 9.1$\times 10^{-4}$\\
\hline
\multirow{15}{*}{Walls} & \multirow{5}{*}{0.0} & 9.0 $\leq \log (\mstar/\msun) < $ 9.5 & 16566 & 0.496 & 0.15 \\
& & 9.5 $\leq \log (\mstar/\msun) < $ 10.0 & 26996 & 0.501 & 0.83 \\
& & 10.0 $\leq \log (\mstar/\msun) < $ 10.5 & 13672 & 0.507 & 9.3$\times 10^{-4}$ \\
& & 10.5 $\leq \log (\mstar/\msun) < $ 11.0 & 5168 & 0.519 & 5.8$\times 10^{-7}$\\
& & $11.0 \leq \log (\mstar/\msun)$ & 1470 & 0.525 & 0.011\\
\cline{2-6}
& \multirow{5}{*}{1.0} & 9.0 $\leq \log (\mstar/\msun) < $ 9.5 & 10734 & 0.502 & 0.15 \\
& & 9.5 $\leq \log (\mstar/\msun) < $ 10.0 & 11230 & 0.506 & 0.038\\
& & 10.0 $\leq \log (\mstar/\msun) < $ 10.5 & 8946 & 0.514 & 2.4$\times 10^{-4}$\\
& & 10.5 $\leq \log (\mstar/\msun) < $ 11.0 & 4498 & 0.52 & 3.1$\times 10^{-5}$\\
& & $11.0 \leq \log (\mstar/\msun)$ & 1256 & 0.527 & 6.9$\times 10^{-4}$\\
\cline{2-6}
& \multirow{5}{*}{2.0} & 9.0 $\leq \log (\mstar/\msun) < $ 9.5 & 7778 & 0.515 & 0.98 \\
& & 9.5 $\leq \log (\mstar/\msun) < $ 10.0 & 5890 & 0.495 & 0.11\\
& & 10.0 $\leq \log (\mstar/\msun) < $ 10.5 & 3188 & 0.502 & 1.4$\times 10^{-5}$\\
& & 10.5 $\leq \log (\mstar/\msun) < $ 11.0 & 1670 & 0.502 & 8.5$\times 10^{-3}$\\
& & $11.0 \leq \log (\mstar/\msun)$ & 682 & 0.494 & 1.4$\times 10^{-3}$\\
\hline
\hline
\end{tabular*}
%%%\begin{tablenotes}
%%%     \item\label{tnote:massBins} stellar mass bins as defined in Section~\ref{section:3Dgalaxies}
%%%     \item\label{tnote:vsig} values for the \hnoagn simulation
%%%%     \item\label{tnote:median} medians of distributions as indicated in Figures\
%%%%              errors represent half width at half maximum of the bootstrap distribution, i.e. the distribution of medians from each of 100 bootstrap samples, fitted by a Gaussian curve
%%%%     \item\label{tnote:mass_grad} Figures
%%%%     \item\label{tnote:type_grad} Figures
%%%%     \item\label{tnote:SF_passive} only galaxies with stellar masses  $\log \, (\msun) \geq 10.46$ are considered
%%%    \end{tablenotes}
\end{threeparttable}
\end{table*}

% ------------------
\begin{table*}
\centering
\begin{threeparttable}
\caption{Number of galaxies, average $\cos \theta$ and the KS probability $p_{\rm KS}$ that the sample is drawn from a uniform distribution for Figure~\ref{fig:spin_fil_vsig}}
\label{tab:vsig}
\begin{tabular*}{0.7\textwidth}{@{\extracolsep{\fill}}lcccc}
\hline
\hline
& \mstar range & $N_{\rm gal}$ & $\left < \cos \theta \right>$ & $p_{\rm KS}$ \\
\hline
\hline
\multirow{4}{*}{Low \vsig} & 9.0 $\leq \log (\mstar/\msun) < $ 9.5 & 11892 & 0.508 & 0.014\\
& 9.5 $\leq \log (\mstar/\msun) < $ 10.0 & 21934 & 0.501 & 0.57\\
& 10.0 $\leq \log (\mstar/\msun) < $ 10.8 & 10682 & 0.499 & 0.2 \\
& $10.8 \leq \log (\mstar/\msun)$ & 1354 & 0.472 & 0.0024\\
%%%%\cline{2-5}
\hline
\multirow{4}{*}{High \vsig} & 9.0 $\leq \log (\mstar/\msun) < $ 9.5 & 4670 & 0.511 & 0.046\\
& 9.5 $\leq \log (\mstar/\msun) < $ 10.0 & 5036 & 0.5 & 0.99\\
& 10.0 $\leq \log (\mstar/\msun) < $ 10.8 & 6544 & 0.491 & 6$\times 10^{-4}$\\
& $10.8 \leq \log (\mstar/\msun)$ & 1484 & 0.485 & 0.08\\
\hline
\hline
\end{tabular*}
\end{threeparttable}
\end{table*}

% ------------------
\begin{table*}
\centering
\begin{threeparttable}
\caption{Number of galaxies, average $\cos \theta$ and the KS probability $p_{\rm KS}$ that the sample is drawn from a uniform distribution for Figure~\ref{fig:spin_fil_ssfr}}
\label{tab:ssfr}
\begin{tabular*}{0.7\textwidth}{@{\extracolsep{\fill}}lcccc}
\hline
\hline
& \mstar range & $N_{\rm gal}$ & $\left < \cos \theta \right>$ & $p_{\rm KS}$ \\
\hline
\hline
\multirow{4}{*}{Low \ssfr} & 9.0 $\leq \log (\mstar/\msun) < $ 9.5 & 2814 & 0.516 & 5.2$\times 10^{-3}$\\
& 9.5 $\leq \log (\mstar/\msun) < $ 10.0 & 5392 & 0.501 & 0.87\\
& 10.0 $\leq \log (\mstar/\msun) < $ 10.8 & 10902 & 0.495 & 0.013 \\
& $10.8 \leq \log (\mstar/\msun)$ & 2232 & 0.476 & 3.5$\times 10^{-4}$\\
%%%%\cline{2-5}
\hline
\multirow{4}{*}{High \ssfr} & 9.0 $\leq \log (\mstar/\msun) < $ 9.5 & 13752 & 0.508 & 0.016\\
& 9.5 $\leq \log (\mstar/\msun) < $ 10.0 & 21604 & 0.501 & 0.76\\
& 10.0 $\leq \log (\mstar/\msun) < $ 10.8 & 6478 & 0.498 & 0.15\\
& $10.8 \leq \log (\mstar/\msun)$ & 698 & 0.469 & 9.3$\times 10^{-3}$\\
\hline
\hline
\end{tabular*}
\end{threeparttable}
\end{table*}

% ------------------
\begin{table*}
\centering
\begin{threeparttable}
\caption{Number of galaxies, average $\cos \theta$ and the KS probability $p_{\rm KS}$ that the sample is drawn from a uniform distribution for Figure~\ref{fig:spin_fil_mHI}}
\label{tab:HI}
\begin{tabular*}{0.7\textwidth}{@{\extracolsep{\fill}}lcccc}
\hline
\hline
& \mstar range & $N_{\rm gal}$ & $\left < \cos \theta \right>$ & $p_{\rm KS}$ \\
\hline
\hline
\multirow{4}{*}{Low HI mass} & 9.0 $\leq \log (\mstar/\msun) < $ 9.5 & 4474 & 0.504 & 0.34\\
& 9.5 $\leq \log (\mstar/\msun) < $ 10.0 & 11732 & 0.497 & 0.34\\
& 10.0 $\leq \log (\mstar/\msun) < $ 10.8 & 7210 & 0.491 & 2$\times 10^{-3}$ \\
& $10.8 \leq \log (\mstar/\msun)$ & 808 & 0.471 & 4$\times 10^{-3}$\\
%%%%\cline{2-5}
\hline
\multirow{4}{*}{High HI mass} & 9.0 $\leq \log (\mstar/\msun) < $ 9.5 & 9656 & 0.509 & 0.01\\
& 9.5 $\leq \log (\mstar/\msun) < $ 10.0 & 12486 & 0.503 & 0.36\\
& 10.0 $\leq \log (\mstar/\msun) < $ 10.8 & 8152 & 0.497 & 0.11\\
& $10.8 \leq \log (\mstar/\msun)$ & 1852 & 0.478 & 5.0$\times 10^{-3}$\\
\hline
\hline
\end{tabular*}
\end{threeparttable}
\end{table*}

% ------------------
\begin{table*}
\centering
\begin{threeparttable}
\caption{Number of galaxies, average $\cos \theta$ and the KS probability $p_{\rm KS}$ that the sample is drawn from a uniform distribution for Figure~\ref{fig:spin_fil_Mh}}
\label{tab:halo}
\begin{tabular*}{0.7\textwidth}{@{\extracolsep{\fill}}lcccc}
\hline
\hline
& \mstar range & $N_{\rm gal}$ & $\left < \cos \theta \right>$ & $p_{\rm KS}$ \\
\hline
\hline
\multirow{3}{*}{Low $M_h$} & 9.0 $\leq \log (\mstar/\msun) < $ 9.5 & 11026 & 0.506 & 0.02\\
& 9.5 $\leq \log (\mstar/\msun) < $ 10.0 & 18046 & 0.501 & 0.44 \\
& 10.0 $\leq \log (\mstar/\msun) < $ 10.5 & 2620 & 0.508 & 0.05\\
%%%%\cline{2-5}
\hline
\multirow{5}{*}{High $M_h$} & 9.0 $\leq \log (\mstar/\msun) < $ 9.5 & 5540 & 0.513 & 0.005 \\
& 9.5 $\leq \log (\mstar/\msun) < $ 10.0 & 8950 & 0.502 & 0.93 \\
& 10.0 $\leq \log (\mstar/\msun) < $ 10.5 & 11052 & 0.497 & 0.036 \\
& 10.5 $\leq \log (\mstar/\msun) < $ 11.0 & 5146 & 0.482 & 8.2$\times 10^{-5}$\\
& $11.0 \leq \log (\mstar/\msun)$ & 1470 & 0.473 & 2.8$\times 10^{-4}$\\
\hline
\hline
\end{tabular*}
\end{threeparttable}
\end{table*}

% ------------------
\begin{table*}
\centering
\begin{threeparttable}
\caption{Number of galaxies, average $\cos \theta$ and the KS probability $p_{\rm KS}$ that the sample is drawn from a uniform distribution for Figure~\ref{fig:spin_fil_density}}
\label{tab:dens}
\begin{tabular*}{0.7\textwidth}{@{\extracolsep{\fill}}lcccc}
\hline
\hline
& \mstar range & $N_{\rm gal}$ & $\left < \cos \theta \right>$ & $p_{\rm KS}$ \\
\hline
\hline
\multirow{4}{*}{Low $\rho$} & 9.0 $\leq \log (\mstar/\msun) < $ 9.5 & 4459 & 0.506 & 0.24\\
& 9.5 $\leq \log (\mstar/\msun) < $ 10.0 & 7756 & 0.505 & 0.08 \\
& 10.0 $\leq \log (\mstar/\msun) < $ 10.5 & 2993 & 0.498 & 0.25\\
& 10.5 $\leq \log (\mstar/\msun) < $ 11.0 & 765 & 0.471 & 0.02 \\
%%%%\cline{2-5}
\hline
\multirow{5}{*}{High $\rho$} & 9.0 $\leq \log (\mstar/\msun) < $ 9.5 & 4133 & 0.514 & 0.001 \\
& 9.5 $\leq \log (\mstar/\msun) < $ 10.0 & 5540 & 0.495 & 0.095 \\
& 10.0 $\leq \log (\mstar/\msun) < $ 10.5 & 3555 & 0.503 & 0.18 \\
& 10.5 $\leq \log (\mstar/\msun) < $ 11.0 & 1841 & 0.494 & 0.39\\
& $11.0 \leq \log (\mstar/\msun)$ & 918 & 0.477 & 0.013\\
\hline
\hline
\end{tabular*}
\end{threeparttable}
\end{table*}

% ------------------
\begin{table*}
\centering
\begin{threeparttable}
\caption{Number of galaxies, average $\cos \theta$ and the KS probability $p_{\rm KS}$ that the sample is drawn from a uniform distribution for Figure~\ref{fig:spin_fil_cenSat}}
\label{tab:cen_sat}
\begin{tabular*}{0.7\textwidth}{@{\extracolsep{\fill}}lcccc}
\hline
\hline
& \mstar range & $N_{\rm gal}$ & $\left < \cos \theta \right>$ & $p_{\rm KS}$ \\
\hline
\hline
\multirow{5}{*}{Centrals} & 9.0 $\leq \log (\mstar/\msun) < $ 9.5 & 10500 & 0.509 & 0.012\\
& 9.5 $\leq \log (\mstar/\msun) < $ 10.0 & 18866 & 0.501 & 0.41\\
& 10.0 $\leq \log (\mstar/\msun) < $ 10.5 & 8988 & 0.495 & 3.2$\times 10^{-3}$ \\
& 10.5 $\leq \log (\mstar/\msun) < $ 10.8 & 2454 & 0.478 & 1.8$\times 10^{-3}$ \\
& $10.8 \leq \log (\mstar/\msun)$ & 2142 & 0.48 & 8.4$\times 10^{-3}$\\
%%%%\cline{2-5}
\hline
\multirow{5}{*}{Satellites} & 9.0 $\leq \log (\mstar/\msun) < $ 9.5 & 6062 & 0.511 & 0.037\\
& 9.5 $\leq \log (\mstar/\msun) < $ 10.0 & 8104 & 0.501 & 0.92\\
& 10.0 $\leq \log (\mstar/\msun) < $ 10.5 & 4574 & 0.506 & 0.021\\
& 10.5 $\leq \log (\mstar/\msun) < $ 10.8 & 1210 & 0.494 & 0.15 \\
& $10.8 \leq \log (\mstar/\msun)$ & 696 & 0.473 & 3.4$\times 10^{-3}$\\
\hline
\hline
\end{tabular*}
\end{threeparttable}
\end{table*}

% ------------------
\begin{table*}
\centering
\begin{threeparttable}
\caption{Number of galaxies, average $\cos \theta$ and the KS probability $p_{\rm KS}$ that the sample is drawn from a uniform distribution for Figure~\ref{fig:galaxy_halo}}
\label{tab:gal_halo}
\begin{tabular*}{0.7\textwidth}{@{\extracolsep{\fill}}lcccc}
\hline
\hline
& \mstar range & $N_{\rm gal}$ & $\left < \cos \theta \right>$ & $p_{\rm KS}$ \\
\hline
\hline
\multirow{5}{*}{Centrals} & 9.0 $\leq \log (\mstar/\msun) < $ 9.5 & 5250 & 0.577 & 2.5$\times 10^{-72}$\\
& 9.5 $\leq \log (\mstar/\msun) < $ 10.0 & 9433 & 0.58 & 2.0$\times 10^{-129}$ \\
& 10.0 $\leq \log (\mstar/\msun) < $ 10.5 & 4494 & 0.599 & 2.8$\times 10^{-90}$ \\
& 10.5 $\leq \log (\mstar/\msun) < $ 10.8 & 1227 & 0.586 & 0.0 \\
& $10.8 \leq \log (\mstar/\msun)$ & 1071 & 0.551 & 3.6$\times 10^{-7}$\\
%%%%\cline{2-5}
\hline
\multirow{5}{*}{Satellites} & 9.0 $\leq \log (\mstar/\msun) < $ 9.5 & 6062 & 0.461 & 8.0$\times 10^{-22}$\\
& 9.5 $\leq \log (\mstar/\msun) < $ 10.0 & 8102 & 0.461 & 4.7$\times 10^{-33}$\\
& 10.0 $\leq \log (\mstar/\msun) < $ 10.5 & 4574 & 0.453 & 6.2$\times 10^{-24}$\\
& 10.5 $\leq \log (\mstar/\msun) < $ 10.8 & 1210 & 0.444 & 1.5$\times 10^{-15}$ \\
& $10.8 \leq \log (\mstar/\msun)$ & 696 & 0.43 & 4.8$\times 10^{-7}$\\
\hline
\hline
\end{tabular*}
\end{threeparttable}
\end{table*}

% ------------------
\begin{table*}
\centering
\begin{threeparttable}
\caption{Number of galaxies, average $\cos \theta$ and the KS probability $p_{\rm KS}$ that the sample is drawn from a uniform distribution for Figure~\ref{fig:halo_filaments}}
\label{tab:halo_gal_fil}
\begin{tabular*}{0.7\textwidth}{@{\extracolsep{\fill}}lcccc}
\hline
\hline
& \mstar range & $N_{\rm gal}$ & $\left < \cos \theta \right>$ & $p_{\rm KS}$ \\
\hline
\hline
\multirow{5}{*}{Centrals} & 9.0 $\leq \log (\mstar/\msun) < $ 9.5 & 10500 & 0.504 & 0.02\\
& 9.5 $\leq \log (\mstar/\msun) < $ 10.0 & 18866 & 0.504 & 0.07 \\
& 10.0 $\leq \log (\mstar/\msun) < $ 10.5 & 8988 & 0.501 & 0.42 \\
& 10.5 $\leq \log (\mstar/\msun) < $ 10.8 & 2454 & 0.491 & 0.22 \\
& $10.8 \leq \log (\mstar/\msun)$ & 2142 & 0.465 & 1.13$\times 10^{-7}$\\
%%%%\cline{2-5}
\hline
\multirow{5}{*}{Satellites} & 9.0 $\leq \log (\mstar/\msun) < $ 9.5 & 3031 & 0.497 & 0.48\\
& 9.5 $\leq \log (\mstar/\msun) < $ 10.0 & 4051 & 0.508 & 0.03\\
& 10.0 $\leq \log (\mstar/\msun) < $ 10.5 & 2287 & 0.499 & 0.39\\
& 10.5 $\leq \log (\mstar/\msun) < $ 10.8 & 605 & 0.495 & 0.47 \\
& $10.8 \leq \log (\mstar/\msun)$ & 348 & 0.489 & 0.56\\
\hline
\hline
\end{tabular*}
\end{threeparttable}
\end{table*}

% ------------------
\begin{table*}
\centering
\begin{threeparttable}
\caption{Number of galaxies, average $\cos \theta$ and the KS probability $p_{\rm KS}$ that the sample is drawn from a uniform distribution for Figure~\ref{fig:halo_filaments_mHaloBins}}
\label{tab:halo_gal_fil_haloBins}
\begin{tabular*}{0.7\textwidth}{@{\extracolsep{\fill}}lcccc}
\hline
\hline
& \mstar range & $N_{\rm gal}$ & $\left < \cos \theta \right>$ & $p_{\rm KS}$ \\
\hline
\hline
\multirow{5}{*}{Centrals} & 10.0 $\leq \log (M_h/\msun) < $ 11.0 & 202 & 0.521 & 0.28\\
& 11.0 $\leq \log (M_h/\msun) < $ 12.0 & 33000 & 0.505 & 0.015 \\
& 12.0 $\leq \log (M_h/\msun) < $ 12.5 & 6434 & 0.501 & 0.13 \\
& 12.5 $\leq \log (M_h/\msun) < $ 13.0 & 2194 & 0.474 & 2.9$\times 10^{-5}$ \\
& $13.0 \leq \log (M_h/\msun)$ & 1120 & 0.44 & 5.2$\times 10^{-11}$\\
%%%%\cline{2-5}
\hline
\multirow{5}{*}{Satellites} & 10.9 $\leq \log (M_h/\msun) < $ 12.0 & 1418 & 0.502 & 0.49\\
& 12.0 $\leq \log (M_h/\msun) < $ 12.5 & 3206 & 0.487 & 1.9$\times 10^{-3}$\\
& 12.5 $\leq \log (M_h/\msun) < $ 13.0 & 3666 & 0.464 & 1.8$\times 10^{-14}$\\
& 13.0 $\leq \log (M_h/\msun) < $ 13.5 & 4192 & 0.432 & 2.0$\times 10^{-44}$ \\
& $13.5 \leq \log (M_h/\msun)$ & 8162 & 0.448 & 2.9$\times 10^{-61}$\\
\hline
\hline
\end{tabular*}
\end{threeparttable}
\end{table*}
%%%%%%%%%%%%%%%%%%%%%%%%%%%%%%%%%%%%%%%%%%%%%%%%%%

% Don't change these lines
\bsp	% typesetting comment
\label{lastpage}
\end{document}